\documentclass[a4paper,11pt]{article}

\usepackage{jheppub} 

\usepackage[T1]{fontenc} 

\title{ Consistently Constrained $SL(N)$ WZWN Models  \\
and Classical Exchange Algebra  }

\author{Shogo Aoyama}
\author{and Katsuyuki Ishii}

\affiliation{Department of Physics, Shizuoka University, \\
 Ohya 836, Shizuoka, Japan}

\abstract{Currents of the $SL(N)$ WZWN model are constrained so that the remaining symmetry is  a symmetry of constrained currents  as well. Such consistency enables us to study the Poisson structure of constrained $SL(N)$ WZWN models properly. We establish the Poisson brackets which satisfy the Jacobi identities owing to the classical Yang-Baxter equation. The Virasoro algebra is shown by using them.  An $SL(N)$ conformal primary is constructed.  It satisfies a quadratic algebra, which might become an exchange algebra by its quantum deformation. }


\arxivnumber{}

\begin{document}

\maketitle

\makeatletter
\def\eqnarray{%
 \stepcounter{equation}%
 \let\@currentlabel=\theequation
 \global\@eqnswtrue
 \global\@eqcnt\z@
 \tabskip\@centering
 \let\\=\@eqncr
 $$\halign to \displaywidth\bgroup\@eqnsel\hskip\@centering
 $\displaystyle\tabskip\z@{##}$&\global\@eqcnt\@ne
 \hfil$\displaystyle{{}##{}}$\hfil
 &\global\@eqcnt\tw@$\displaystyle\tabskip\z@{##}$\hfil
 \tabskip\@centering&\llap{##}\tabskip\z@\cr}
\makeatother

\renewcommand{\theequation}{\thesection.\arabic{equation}}

\section{Introduction}

The $2$-dimensional WZWN model is considered as a natural generalization of the Nambu-Goto string. Having the Virasoro-Kac-Moody symmetry it provides an arena for interesting studies of the CFT. Among them the constrained $SL(2)$ WZWN model  was extensively studied in the late '80s in connection with the non-critical string. The studies stimulated much interest in close relationships of various subjects  like the Liouville theory, light-cone gauge or geometrical formulation of the string, Poisson structure, quantum group etc..  For a summary of the arguments and references  the reader may refer to \cite{Ao}. For  constrained $SL(N)$ WZWN models with $N\ge 3$  those subjects  were  not understood so fully as for the $SL(2)$ model.  
The case of $SL(3)$ was initiated by Polyakov\cite{Po} and studied by many people with interest in finding $W$ algebra as an extension of the Virasoro algebra. In particular in \cite{Be} they gave a rather comprehensive arguments for the subject. Studying  the $SL(N)$ WZWN model  by equivalent free field theories 
 had been already a well-known approach in the literature\cite{Wa}.  In \cite{Be} they succeeded in formulating  a constrained $SL(N)$ WZWN model by reducing the phase space of the unconstrained model in this free field approach. Quantization was successfully done. The reader may refer to \cite{Bo} for a more extensive  review on the subject. 
But they did not give  a proper account on Poisson structure of the constrained $SL(N)$ WZWN model, and consequently quantum group structure was not discussed.
 Since then the subject was left intact to the authors' knowledges. It is worth  reviving attention on the subject in view of 
 the fact that  non-linear $\sigma$-models with $PSL(4|4)$ or  $PSL(2,2|4)$\cite{Bei,Bek2}  might bring us  new aspects of the string/QCD duality\cite{Ma}.

 This paper is organized as follows. In section 2  we give a brief summary about the known results of  constrained $SL(2)$ WZWN models to the extent of our interest.  The constrained $SL(2)$ WZWN model is  formulated by gauging the ordinary  WZWN model. 
In section 3 we extend the model to the case of $SL(3)$. An important point is that gauge-coupled currents are constrained consistently so that the remaining symmetry after gauge-fixing is a symmetry of themselves as well. 
 In this section we study the case where the gauge-fixed symmetry is 
{\it irreducible}, i.e., the symmetry is realized by the irreducible coset space $SL(3)/\{SL(2)\otimes U(1)\}$.\footnote{The coset space $G/H$ is said to be {\it irreducible} when the coset generators belong to an irreducible representation of the homogeneous group $H$\cite{AAo}.} Owing to the symmetry of the constrained currents we set up Poisson brackets consistently. The classical Yang-Baxter equation plays a key role for  consistency of  Poisson brackets. By using them we  derive the Virasoro algebra for an improved energy-momentum tensor.  We also find an $SL(3)$ conformal primary and show it to satisfy a  quadratic algebra. It may be called classical exchange algebra since it may be deformed to a quantum exchange algebra obeying the Yang-Baxter equation. The arguments in section 3 can be straightforwardly generalized to the case of irreducibly constrained $SL(N)$ WZWN models with $N\ge 4$. They are given in Sections 4.  We also study the conformal transformation of the constrained currents. They should transform as primaries of weight $0$ with respect to an improved energy-momentum tensor.  Otherwise it would not be possible to constrain the $SL(N)$ WZWN model consistently. We give a general proof for this property of the constrained currents.
 Section 5 is devoted to extend the arguments to reducibly constrained WZWN models. As an example we study the case 
where the gauge-fixed symmetry is realized by the reducible coset space $SL(N)/U(1)^{N-1}$. We show that  the whole arguments from section 2 to 4 go through for the reducible case as well.

 $SL(N)$ conformal primaries  were discussed by the free field realization\cite{Wa, Ito,Bek} of the current algebra of the WZWN model in \cite{PRY,Ao2}. 
We present the arguments of \cite{Ao2} in appendix A. We think that it is worth doing since  the latter approach might bring us more insight for quantum consideration of the $SL(N)$ conformal primary of this paper.

\vspace{1cm}

\section{Summary of the constrained $SL(2)$ WZWN model}
\setcounter{equation}{0}

We give a review on  the constrained $SL(2)$ WZWN model and the related subjects referring to \cite{Ao}. 
The constrained $SL(2)$ WZWN model is given by 
\begin{eqnarray}
S=-{k\over 4\pi}S_{WZWN}-{k\over 2\pi}\int d^2x {\rm Tr}[A_-(g^{-1}\partial_+g-e)],    \label{action} 
\end{eqnarray}
with 
\begin{eqnarray}
A_-=\left(
\begin{array}{cc}
0 & a_- \\
0 & 0 
\end{array}
\right), \quad\quad\quad e:{\rm const.}\ 2\times 2\ {\rm matrix}.  \nonumber
\end{eqnarray}
The equation of motion for the gauge field $A_-$ provides the constraint for the $SL(2)$ current
\begin{eqnarray}
J_+\propto    {\rm Tr}\Big[ g^{-1}\partial_+g\left(
\begin{array}{cc}
0 & 1 \\
0 & 0 
\end{array}
\right)
\Big]={\rm const.}.    \label{constraint}
\end{eqnarray}
A group element can be parametrized as
\begin{eqnarray}
g=\left(
\begin{array}{cc}
1 &  0 \\
G &  1 
\end{array}
\right)
\left(
\begin{array}{cc}
\lambda & 0 \\
0 & {1\over\lambda} 
\end{array}
\right)
\left(
\begin{array}{cc}
1 & F \\
0 & 1 
\end{array}
\right)\equiv g_Lg_0g_R.    \label{parametrization}    
\end{eqnarray}
We choose the unitary gauge $F=0$ by fixing  $A_-$. Then the constraint (\ref{constraint})
 takes the form $\lambda^2 \partial_+G={\rm const.}$, owing to which the action (\ref{action}) becomes the one of the $2d$ effective gravity in the geometrical formulation\cite{AS}
\begin{eqnarray}
S= -{k\over 8\pi}\int d^2x {\partial_-G\over \partial_+G}\Bigg[{\partial_+^3G\over \partial_+G} - 2\Big({\partial_+^2G\over \partial_+G}\Big)^2\Bigg].
\label{gravity}
\end{eqnarray}
This theory has remarkable properties. First of all, 
under a diffeomorphism transformation in the right-moving sector 
\begin{eqnarray}
\delta G= \eta(x^+)\partial_+G,   \label{diffeo} 
\end{eqnarray}
it is invariant transforming as 
$$
\delta S= {1\over 2\pi}\int d^2x\ \eta(x^+)\partial_-T_{++} 
$$
with 
\begin{eqnarray}
T_{++}={k\over 2}\Bigg[{\partial_+^3G\over \partial_+G} - {3\over 2}\Big({\partial_+^2G\over \partial_+G}\Big)^2\Bigg]\quad\quad ({\rm Schwarzian\  derivative}).  \label{Schwarz}
\end{eqnarray}
Secondly it is also invariant under the $SL(2)$ transformation in the left-moving sector 
\begin{eqnarray}
\delta G= {\epsilon_R(x^-) \over \sqrt 2}- \epsilon_0(x^-)G-{\epsilon_L(x^-)\over \sqrt 2}G^2\equiv \epsilon_A\delta^AG.
  \label{Killing}
\end{eqnarray}
Here $\delta^A G$ are a set of the Killing vectors which non-linearly realize the 
$SL(2)$ Lie algebra. The group elements $g_L,g_0,g_R$ in the parametrization (\ref{parametrization})  are generated by
\begin{eqnarray}
T_L={1\over \sqrt 2}\left(
\begin{array}{cc}
0 &\ \ 0 \\
1 &\ \  0 
\end{array}
\right),\quad\quad 
T_0={1\over  2}\left(
\begin{array}{cc}
1 &\ \ 0 \\
0 & -1 
\end{array}
\right),\quad\quad 
T_R={1\over \sqrt 2}\left(
\begin{array}{cc}
0 &\  \ 1 \\
0 &\ \  0 
\end{array}
\right).   \label{T}
\end{eqnarray}
By these generators the quadratic Casimir takes the form \footnote{ $e^{\epsilon\cdot T}\in SL(2)$ with $\epsilon$ real parameters. But $e^{\epsilon\cdot T}\in SU(2)$ with $\epsilon$ pure imaginary ones. The normalization (\ref{T}) was chosen so that the quadratic Casimir $C_{adj}$ is $-2$.  }
$$
-C=T_RT_L+T_LT_R+T_0^2 \equiv T\cdot T.
$$

We may set  a Poisson bracket for the theory (\ref{gravity}) as
\begin{eqnarray}
\{G(x)\mathop{,}^\otimes G(y)\}=-{\pi\over k}\Big[\epsilon(x-y)(G(x)-G(y))^2
 +G(x)^2-G(y)^2\Big]    \label{Poi1}
\end{eqnarray}
for $x^-=y^-$\cite{AS2}. It can be shown that this satisfies the Jacobi identity.  This Poisson bracket correctly reproduces the respective diffeomorphism for $G$ and the energy-momentum tensor (\ref{Schwarz}) as
\begin{eqnarray}
   {1\over 2\pi}  \int dx\ \eta(x)\{T_{++}(x)\mathop{,}^\otimes G(y)\}&=&\eta(y)\partial_yG(y),  \nonumber\\
{1\over 2\pi}\int dx\ \eta(x)\{T_{++}(x)\mathop{,}^\otimes T_{++}(y)\}&=&\eta(y)T'_{++}(y)+2\eta'(y)T_{++}(y)-{k\over 2}\eta'''(y), \nonumber
\end{eqnarray}
for $x_-=y_-$. We now define the quantity
\begin{eqnarray}
\Psi= {1\over \sqrt {G'}}{1 \choose  G}.   \label{primary2}
\end{eqnarray}
This deserves to be called  $SL(2)$ conformal primary for the following reasons. Firstly 
it linearly transforms as a $2$-dimensional  $SL(2)$ spinor, i.e.,
\begin{eqnarray}
\delta \Psi= \epsilon\cdot T \Psi   \nonumber
\end{eqnarray}
by the transformation (\ref{Killing}). Secondly it transforms 
\begin{eqnarray}
\int dx\ \eta(x)\{T_{++}(x)\mathop{,}^\otimes \Psi(y)\}=\eta(y)\Psi'(y) -{1\over 2}\eta'(y)\Psi(y) \nonumber
\end{eqnarray}
with  weight $-{1\over 2}$ in the right-moving sector.   Using the Poisson bracket we can show that the primary satisfies a quadratic algebra
\begin{eqnarray}
\{\Psi(x)\mathop{,}^\otimes \Psi(y)\}={2\pi\over k}[\theta(x-y)r^+ + \theta(y-x) r^-]\Psi(x)\otimes\Psi(y).   \label{CEA1}
\end{eqnarray}
Here $r^+$ and $r^-$ are the $r$-matrices
\begin{eqnarray}
r^+=2T_R\otimes T_L +T_0\otimes T_0,\quad\quad 
r^-=-2T_L\otimes T_R -T_0\otimes T_0,    \nonumber
\end{eqnarray}
 which satisfy the classical Yang-Baxter equation. We may think of the quadratic algebra (\ref{CEA1}) as a classical version of a quantum exchange algebra
\begin{eqnarray}
\Psi(x)^a\otimes \Psi(y)^b= [\theta(x-y)R^+ + \theta(y-x)R^-]^{ab}_{cd}\Psi^d(y)\otimes \Psi^c(x),  \nonumber
\end{eqnarray}
by deforming the $r$-matrices to 
$$
R^{\pm}= 1\otimes 1 +hr^{\pm}+ O(h^2)      
$$
with  $h={2\pi\over k}$.

\vspace{1cm}

\section{A constrained $SL(3)$ WZWN model}

\subsection{Symmetries of  constrained currents}

As a generalization of the action (\ref{action}) we propose a constrained $SL(3)$ WZWN model given by 
\begin{eqnarray}
S=-{k\over 4\pi}S_{WZWN}-{k\over 2\pi}\int d^2x {\rm Tr}[A_-(g^{-1}\partial_+g-e)],    \label{action2}
\end{eqnarray}
with 
\begin{eqnarray}
g\in SL(3),\quad\quad A_-=\left(
\begin{array}{ccc}
\hspace{0.1cm} 0 &\hspace{0.15cm} a_-^1 & a_-^2 \\
\hspace{0.1cm} 0 &\hspace{0.15cm} 0 & 0 \\
\hspace{0.1cm} 0 &\hspace{0.15cm} 0 & 0 
\end{array}
\right), \quad\quad e:{\rm const.}\ 3\times 3\ {\rm matrix}.  \label{gauge}
\end{eqnarray} 
The form of the gauge field suggests to take a parametrization  based  on the subgroup $SL(2)\otimes U(1)$ such as
$
g=g_Lg_0g_R
$
with 
\begin{eqnarray}
g_L=\left(
\begin{array}{ccc}
1 & 0 &\hspace{0.15cm} 0 \\
G_1 & 1 &\hspace{0.15cm} 0 \\
G_2 & 0 &\hspace{0.15cm} 1 
\end{array}\right), \quad\quad 
g_0=\left(
\begin{array}{ccc}
{1\over\Delta} &\hspace{0.1cm} 0 &\hspace{0.1cm} 0 \\
0 &\hspace{0.1cm} \lambda &\hspace{0.1cm} F \\
0 &\hspace{0.1cm} G &\hspace{0.1cm} \mu 
\end{array}\right), \quad\quad 
g_R=\left(
\begin{array}{ccc}
\hspace{0.1cm}1 &\hspace{0.05cm} F^1 &\hspace{-0.1cm} F^2 \\
\hspace{0.1cm}0 &\hspace{0.05cm} 1 &\hspace{-0.1cm} 0 \\
\hspace{0.1cm}0 &\hspace{0.05cm} 0 &\hspace{-0.1cm} 1 
\end{array}\right),  \label{para1}
\end{eqnarray}
in which $\Delta=\lambda\mu-FG$. It will become clear soon later why we start  by taking this parametrization as well as the form of (\ref{gauge}). 
We choose the unitary gauge $F^1=F^2=0$. Then  the equation of motion for $A_-$  yields the constraints
\begin{eqnarray}
J_{+1} &=&  {\rm Tr}\Big[ g^{-1}\partial_+g\left(
\begin{array}{ccc}
 0 & \ \  1 &\ \ 0 \\
 0 &\ \ 0 &\ \ 0 \\
 0 &\ \ 0 &\ \  0
\end{array}
\right)
\Big]={1\over \Delta^2}(\mu\partial_+G_1-F\partial_+G_2)={\rm const.},     \nonumber \\
J_{+2} &=&  {\rm Tr}\Big[ g^{-1}\partial_+g\left(
\begin{array}{ccc}
0 & \ \  0 &\ \ 1 \\
 0 &\ \ 0 &\ \  0 \\
 0 &\ \ 0 &\ \  0
\end{array}
\right)
\Big] = {1\over \Delta^2}(-G\partial_+G_1+\lambda\partial_+ G_2)={\rm const.}.      \label{current}
\end{eqnarray}
We look for a transformation for the group variables ${\cal G}^I=(G_1,G_2, F,G,\lambda,\mu)$ which leaves the constraints invariant, or equivalently the unitary gauge fixed. It can be constructed as follows. Consider the quantity $g({\cal G})=g_L(G_1,G_2)g_0(\lambda,F,G,\mu)$.  
By a left multiplication of $e^{\epsilon\cdot T}\in SL(3)$ we find the relation
\begin{eqnarray}
g({\cal G})\longrightarrow e^{\epsilon\cdot T}g({\cal G}) U_R^{-1}=g({\cal G}'),   \label{SL(3)}
\end{eqnarray}
appropriately choosing a compensator $U_R$ as
\begin{eqnarray}
U_R^{-1}=\left(
\begin{array}{ccc}
1 & \ \  * &\ \ * \\
 0 &\ \ 1 &\ \  0 \\
 0 &\ \ 0 &\ \  1
\end{array}  
\right)\equiv 1- u.        \nonumber
\end{eqnarray}
This defines the symmetry transformation of the group variables ${\cal G}^I$ to ${\cal G}'^I$, which we have looked for. We postpone calculation of their concrete forms to the end of this section, but  discuss the symmetry by the transformation at first. If $e^{\epsilon\cdot T}$ depends on $x^-$ alone, the  transformation (\ref{SL(3)}) induces 
\begin{eqnarray}
\delta(g^{-1}\partial_+g)=-\partial_+u-[g^{-1}\partial_+g,u]  \label{property}
\end{eqnarray}
for infinitesimal parameters $\epsilon$. 
 Then it follows that  the right-moving currents (\ref{current}) are invariant by the transformation. This invariance guarantees that imposing the constraints (\ref{current}) does not restrict the gauge-fixed symmetry  by transformation (\ref{SL(3)}) furthermore. This property makes the argument easy. This is the reason why we started by gauging as (\ref{gauge}). 
The action (\ref{action2}) is also invariant since we have 
$$
\delta S= -{k\over 2\pi}\int d^2x {\rm Tr} [\epsilon\cdot T\partial_+(\partial_-gg^{-1})
 - u\partial_-(g^{-1}\partial_+g)]=0,
$$
owing to $\partial_+\epsilon\cdot T=0$ and the constraints (\ref{current}). 
Here we have used 
\begin{eqnarray}
\delta S_{WZWN}= -2\int d^2x{\rm Tr}[(\partial_-gg^{-1})\partial_+(\delta gg^{-1})]= -2\int d^2x{\rm Tr}[(g^{-1}\partial_+g)\partial_-(g^{-1}\delta g)].\quad\quad\ \  \label{WZtrans}
\end{eqnarray}
 The invariance implies also that the  left-moving currents is conserved 
as
\begin{eqnarray}
\partial_+(\partial_-gg^{-1})=0.       \nonumber
\end{eqnarray}
Applying  the formula  (\ref{WZtrans}) to diffeomorphism $\delta g=\eta(x^+)\partial_+ g$ leads us to find 
$$
\delta S_{WZWN}= -\int d^2x\ \eta(x^+)\partial_-{\rm Tr}[(g^{-1}\partial_+g)^2]
$$
with the use of (\ref{WZtrans}). 
 We define the modified energy-momentum tensor in the right-moving sector as
\begin{eqnarray}
T_{++}&=& k\Bigg({1\over 2}{\rm Tr}(g^{-1}\partial_+g)^2+\partial_+{\rm Tr}[(g^{-1}\partial_+g)T^{U(1)}]\Bigg),   \label{modT}
\end{eqnarray}
with $T^{U(1)}$  satisfying 
$
 [u, T^{U(1)}]=-u \label{u}
$, i.e.,
\begin{eqnarray}
T^{U(1)}={1\over 3}\left(
\begin{array}{ccc}
\ 2\ &   0 & 0 \\
\ 0\ & -1 & 0 \\
\ 0\ & 0 & -1 
\end{array}\right).
  \nonumber
\end{eqnarray}
The modified energy-momentum tensor turns out to be invariant by the symmetry transformation (\ref{SL(3)})  as 
\begin{eqnarray}
\delta T_{++}= -k{\rm Tr}[(g^{-1}\partial_+g)\partial_+u] -k\partial_+{\rm Tr}((g^{-1}\partial_+g)[u,T^{U(1)}])=0. \label{ET}
\end{eqnarray}
Here we have used  (\ref{property}) and the constraints (\ref{current}).
 
Finally we show the transformation law of ${\cal G}^I=(G_1,G_2, F,G,\lambda,\mu)$ from (\ref{SL(3)}). We fix the normalization of the Lie algebra  of $SL(3)$  
$$
[T^A,T^B]=f^{AB}_{\ \ \ C}T^C,
$$
 by taking the quadratic Casimir in the form 
\begin{eqnarray}
-C=T\cdot T&\equiv& t_{AB}T^AT^B   \label{norm}  \\
&\equiv& T_{R1} T_L^1+T_{R2}T_L^2+T_{R3}T_L^3+
T_L^1 T_{R1}+T_L^2 T_{R2}+T_L^3 T_{R3}+
T^QT^Q+T^YT^Y. \quad\quad  \nonumber
\end{eqnarray}
Here $T_L^1,T_L^2(T_{R1},T_{R2})$ are the generators of $g_L(g_R)$ in (\ref{para1}), while the remaining $T$s are those of $g_0 \in SL(2)\otimes U(1)$. To be explicit they are
\begin{eqnarray}
&\ & T^Q={1\over  2}\left(
\begin{array}{ccc}
\ 0\ & \ 0 &\ 0  \\
\ 0\ & \ 1  &\ 0 \\
\ 0\ & \ 0 &\ -1
\end{array}
\right), \quad\quad  
T^Y={1\over2\sqrt 3} \left(
\begin{array}{ccc}
\ 2\ &   0 &\  0 \\
\ 0\ & -1 &\  0 \\
\ 0\ & 0 & -1
\end{array}
\right)={\sqrt 3\over 2}T^{U(1)},\quad      \nonumber\\
&\quad&  T_L^1={1\over \sqrt 2}\left(
\begin{array}{ccc}
\ 0\ & \   0\ &\  0\ \\
\ 1\ &\ 0\ &\  0\ \\
\ 0\ &\ 0\ &\   0\
\end{array}
\right), \quad\ \ 
 T_{R1}={1\over \sqrt 2}\left(
\begin{array}{ccc}
\ 0\ & \   1\ &\  0\ \\
\ 0\ &\ 0\ &\  0\ \\
\ 0\ &\ 0\ &\   0\
\end{array}
\right), \quad {\rm etc.}, 
    \label{gene}
\end{eqnarray}
so that the quadratic Casimir $C_{adj}$ is $-3$.

We study the transformation  (\ref{SL(3)}) at  two steps  such as
\begin{eqnarray}
g_L(G_1,G_2)&\longrightarrow& e^{\epsilon\cdot T}g_L(G_1,G_2) e^{-\rho\cdot \hat H}=g_L(G'_1,G'_2),   \label{1st} \\
g_0(\lambda,F,G,\mu)&\longrightarrow& e^{\rho\cdot \hat H}g_0(\lambda,F,G,\mu)U_R^{-1}=g_0(\lambda',F',G',\mu'),  \label{2nd}
\end{eqnarray}
in which 
\begin{eqnarray}
\epsilon\cdot T &=&\epsilon_{R1} T_L^1+\epsilon_{R2}T_L^2+\epsilon_{R3}T_L^3+
\epsilon_L^1T_{R1}+\epsilon_L^2 T_{R2}+\epsilon_L^3 T_{R3}+
\epsilon_QT^Q+\epsilon_YT^Y\equiv t_{AB}\epsilon^AT^B, \nonumber   \\
\rho\cdot \hat H &=& \rho_{R3}T_L^3+
\rho_L^1T_{R1}+\rho_L^2 T_{R2}+\rho_L^3 T_{R3}+
\rho_QT^Q+\rho_YT^Y.    \nonumber
\end{eqnarray}
The first step relation gives the transformation law for $G_1$ and $G_2$ 
\begin{eqnarray}
\delta G_1 &=&  {\epsilon_{R1}\over \sqrt 2}+{\epsilon_L^3\over \sqrt 2}G_2-\Big({\sqrt 3\over 2}\epsilon_Y-{\epsilon_Q\over  2}+
 {\epsilon_L^1\over \sqrt 2}G_1+{\epsilon_L^2\over \sqrt 2}G_2\Big)G_1\equiv t_{AB}\epsilon^A\delta^BG_1,    \nonumber \\
\delta G_2 &=& {\epsilon_{R2}\over \sqrt 2}+{\epsilon_{R3}\over \sqrt 2}G_1-\Big({\sqrt 3\over 2}\epsilon_Y+{\epsilon_Q\over  2}
 +{\epsilon_L^1\over \sqrt 2}G_1+{\epsilon_L^2\over \sqrt 2}G_2\Big)G_2 \equiv t_{AB}\epsilon^A\delta^BG_2,         \label{G-trans}
\end{eqnarray}
together with the infinitesimal compensator 
\begin{eqnarray}
\rho\cdot \hat H&=&
\left(
\begin{array}{ccc}
{\epsilon_Y\over \sqrt 3}+{\epsilon_L^1\over \sqrt 2}G_1+{\epsilon_L^2\over \sqrt 2}G_2 & {\epsilon_L^1\over \sqrt 2} & 
  {\epsilon_L^2\over \sqrt 2} \\
0  &{\epsilon_Q\over  2}-{\epsilon_Y\over 2\sqrt 3}-{\epsilon_L^1\over \sqrt 2}G_1 & {\epsilon_L^3\over \sqrt 2}-{\epsilon_L^2\over \sqrt 2}G_1 \\
0 & {\epsilon_R^3\over \sqrt 2}-{\epsilon_L^1\over \sqrt 2}G_2 & -{\epsilon_Q\over 2}-{\epsilon_Y\over 2\sqrt 3}-{\epsilon_L^2\over \sqrt 2}G_2  
\end{array}
\right). \quad
  \label{trans}
\end{eqnarray}
The  second step relation  gives the transformation law for $F,G, \lambda,\mu$ 
\begin{eqnarray}
\delta\lambda &=& ({\epsilon_Q\over 2}-{\epsilon_Y\over 2\sqrt 3}-{\epsilon_L^1\over \sqrt 2}G_1)\lambda
    + ({\epsilon_L^3\over \sqrt 2}-{\epsilon_L^2\over \sqrt 2}G_1)G \equiv t_{AB}\epsilon^A\delta^B\lambda,
\nonumber\\
\delta F &=& ({\epsilon_Q\over 2}-{\epsilon_Y\over 2\sqrt 3}-{\epsilon_L^1\over \sqrt 2}G_1)F
    + ({\epsilon_L^3\over \sqrt 2}-{\epsilon_L^2\over \sqrt 2}G_1)\mu \equiv t_{AB}\epsilon^A\delta^BF,
\nonumber\\
\delta G &=&({\epsilon_R^3\over \sqrt 2}-{\epsilon_L^1\over \sqrt 2}G_2)\lambda
   -({\epsilon_Q\over 2}+{\epsilon_Y\over 2\sqrt 3}+{\epsilon_L^2\over \sqrt 2}G_2)G
\equiv t_{AB}\epsilon^A\delta^BG,
     \nonumber\\
\delta \mu &=&({\epsilon_R^3\over \sqrt 2}-{\epsilon_L^1\over \sqrt 2}G_2)F
   -({\epsilon_Q\over 2}+{\epsilon_Y\over 2\sqrt 3}+{\epsilon_L^2\over \sqrt 2}G_2)\mu
\equiv t_{AB}\epsilon^A\delta^B\mu,    \label{lambda}
\end{eqnarray}
together with 
\begin{eqnarray}
u=
\left(
\begin{array}{ccc}
\hspace{0.5cm}0\hspace{1cm} & {\Delta\over \sqrt 2}(\epsilon_L^1\lambda+\epsilon_L^2G) & 
 {\Delta\over \sqrt 2}(\epsilon_L^1F+\epsilon_L^2\mu)  \\
\hspace{0.5cm}0\hspace{1cm}  &   0  & 0 \\
\hspace{0.5cm}0\hspace{1cm} & 0 & 0 
\end{array}
\right). 
  \label{trans2}
\end{eqnarray}
$G_1$ and $G_2$ are  coordinates of the coset space $SL(3)/\{SL(2)\otimes U(1)\}$. The transformations (\ref{G-trans}) and (\ref{lambda}) give the Killing vectors. But $\lambda,F,G,\mu$ are merely auxiliary coordinates for the coset space.  The coset space is irreducible because the coordinates $G_1$ and $G_2$ belong to the fundamental representaion of the homogeneous group $SL(2)$. 
In the beginning of this subsection we have rather formally shown that the right-moving currents
$J_{+1}$ and $J_{+2}$ are invariant by the transformation (\ref{SL(3)}). Here it can be directly checked by varying the expression  (\ref{current}) by these Killing vectors.

\subsection{Poisson brackets and the Virasoro algebra}

We shall set up Poisson brackets for the group variables ${\cal G}^I=(G_1,G_2, F,G,\lambda,\mu)$. The guiding principle to do this is that they satisfy the Jacobi identities and are able to reproduce the Virasoro algebra for the energy-momentum tensor (\ref{modT}). We shall show that they are given by 
\begin{eqnarray}
&\ & \{{\cal G}^I(x)\mathop{,}^\otimes {\cal G}^J(y)\}   \nonumber \\
&\ &\quad\quad ={2\pi\over k}\Big[\theta(x-y)t_{AB}^+ \delta^A{\cal G}^I(x)\otimes \delta^B{\cal G}^J(y) -\theta(y-x)t_{AB}^+ \delta^A{\cal G}^J(y)\otimes \delta^B{\cal G}^I(x)\Big]\quad\     \label{Poisson}
\end{eqnarray}
at $x^-=y^-$. The notation is as follows. $\theta(x)$ is the step function. $\delta^A {\cal G}^I(x)$ are given by (\ref{G-trans}) and (\ref{lambda}), which are the Killing vectors of the coset space $SL(3)/\{SL(2)\otimes U(1)\}$. More correctly they should be written as $\delta {\cal G}^I({\cal G}(x))$, but the dependence of ${\cal G}^I(x)$ was omitted to avoid an unnecessary complication. The quantity $t_{AB}^+$ is the most crucial in  our arguments. It is the modified Killing metric which defines the classical $r$-matrices
 as
\begin{eqnarray}
r^{\pm}&=&T_{R1}\otimes T_L^1+T_{R2}\otimes T_L^2+T_{R3}\otimes T_L^3-
T_L^1\otimes T_{R1}-T_L^2\otimes T_{R2}-T_L^3\otimes T_{R3} \pm t_{AB}T^A\otimes T^B  \nonumber\\
  &\equiv&  t_{AB}^{\pm}T^A\otimes T^B.    \nonumber
\end{eqnarray}
For $r^+$ it reads
$$
r^{+}=2 T_{R1}\otimes T_L^1+2T_{R2}\otimes T_L^2+2T_{R3}\otimes T_L^3 
    +T^Q\otimes T^Q+ T^Y\otimes T^Y.
$$
They satisfy the classical Yang-Baxter equation\cite{J}
\begin{eqnarray}
[r_{12}^{\pm},r_{13}^{\pm}]+[r_{12}^{\pm},r_{23}^{\pm}]+[r_{13}^{\pm},r_{23}^{\pm}]=0.  \label{CYB}
\end{eqnarray}
Here 
$$
r_{12}^{\pm}=t_{AB}^{\pm}T^A\otimes T^B\otimes 1,\quad 
r_{13}^{\pm}=t_{AB}^{\pm}T^A\otimes 1\otimes T^B ,\quad
r_{23}^{\pm}=t_{AB}^{\pm}1\otimes T^A\otimes T^B.
$$
Owing to the classical Yang-Baxter equation the Poisson brackets (\ref{Poisson}) indeed satisfy the Jacobi identities as follows.  Choose any three group variables from ${\cal G}^I$, say  $X,Y,Z$. Assume that $x>y>z$.  After a little algebra we find 
\begin{eqnarray}
&\ & \{X(x)\mathop{,}^\otimes \{Y(y)\mathop{,}^\otimes Z(z)\}\}+
\{Y(y)\mathop{,}^\otimes \{Z(z)\mathop{,}^\otimes X(x)\}\}  +
\{Z(z)\mathop{,}^\otimes \{X(x)\mathop{,}^\otimes Y(y)\}\} 
 \nonumber\\
&\ &\quad\quad \propto -t_{AB}^+t_{CD}^+\Big(
\delta^{[C}\delta^{A]}X(x)\otimes\delta^DY(y)\otimes\delta^BZ(z)  
+ \delta^CX(x)\otimes\delta^{[D}\delta^{A]}Y(y)\otimes\delta^BZ(z) \nonumber\\
&\ &\hspace{7.5cm}+ \delta^CX(x)\otimes\delta^AY(y)\otimes\delta^{[D}\delta^{B]}Z(z) \Big).  \label{XYZ}
\end{eqnarray}
By the construction it is obvious that the Killing vectors $\delta^A X, \delta^A Y, \delta^A Z $, say  $\delta^A {\cal G}$, satisfy the Lie algebra of $SL(3)$ 
\begin{eqnarray}
\delta^{[B}\delta^{A]}{\cal G}(x)=f^{AB}_{\ \ \ C}\delta^C{\cal G}(x).  \label{Lie1}
\end{eqnarray}
Consequently we have 
\begin{eqnarray}
\delta^C{\cal G}(x)={1\over C_{adj}}f_{AB}^{\ \ \ C}\delta^{[A}\delta^{B]}{\cal G}(x),   \label{Lie2}
\end{eqnarray}
by our normalization $-f_{AB}^{\ \ \ C}f^{AB}_{\ \ \ D}=C_{adj}\delta_D^C$ with
 $C_{adj}=-3$. Putting (\ref{Lie1}) and (\ref{Lie2}) in the {\it r.h.s.} of the last line of (\ref{XYZ}) yields 
 the {\it l.h.s.} of the classical Yang-Baxter equation (\ref{CYB}) in the adjoint representation. Therefore the Jacobi identities for the Poisson brackets are satisfied. Hitherto we argued by assuming  $x>y>z$. The arguments hold even if we take other orders for $x,y,z$. We would like to remark that the Poisson bracket (\ref{Poi1}) for the case of $SL(2)$ can be obtained by applying the general formula (\ref{Poisson}) with (\ref{Killing}).
 
Now we shall show how to obtain the Virasoro algebra  by means of the Poisson brackets (\ref{Poisson}). 
The Poisson bracket of our interest reads 
\begin{eqnarray}
&\ & \{T_{++}(x) \mathop{,}^\otimes T_{++}(y)\}      \nonumber\\
&\ &\quad\quad\quad = k\Bigg({\rm Tr}[(g^{-1}\partial_x g)\{g^{-1}\partial_x g\mathop{,}^\otimes T_{++}(y)\} ]+{2\over \sqrt 3}
\partial_x{\rm Tr}[\{g^{-1}\partial_x g\mathop{,}^\otimes T_{++}(y)\}T^Y]\Bigg)
\nonumber
\end{eqnarray}
at $x^-=y^-$.  With the help of the formula for a generic variation
\begin{eqnarray}
\delta(g^{-1}\partial_xg)=g^{-1}\partial_x(\delta g g^{-1})g,  \label{variation}\end{eqnarray}
it becomes 
\begin{eqnarray}
&\ &\{T_{++}(x) \mathop{,}^\otimes T_{++}(y)\}     \nonumber\\
&\ & =k\Bigg({\rm Tr}[(\partial_x gg^{-1})\partial_x(\{g\mathop{,}^\otimes T_{++}(y)\}g^{-1})]   
+ {2\over \sqrt 3}\partial_x{\rm Tr}[\partial_x(\{g\mathop{,}^\otimes T_{++}(y)\}g^{-1})gT^Yg^{-1}]\Bigg). \quad\quad\ \ \ \label{TT}
\end{eqnarray}
Here keep in mind the $x$-dependence of $g$ which  was omitted for simplicity. 
We further calculate the Poisson bracket  $\displaystyle{\{g\mathop{,}^\otimes T_{++}(y)\}}$ in the {\it r.h.s.}. It can be done by the same procedure as for obtaining  (\ref{TT}). Then  we have 
\begin{eqnarray}
&\ & \{g(x)\mathop{,}^\otimes T_{++}(y)\}  \nonumber\\
&\ &\quad\quad 
= k\Bigg({\rm Tr}[\partial_y(\{g(x)\mathop{,}^\otimes g\}g^{-1})
(\partial_y gg^{-1})]+{2\over \sqrt 3}\partial_y{\rm Tr}[\partial_y(\{g(x)\mathop{,}^\otimes g\}g^{-1})gT^Yg^{-1}]\Bigg), \quad\quad \label{gg}
\end{eqnarray}
omitting the $y$-dependence of $g$  for simplicity   this time. Here we understand  also that ${\rm Tr}$ in the {\it r.h.s.} acts on $g(y)$ and does not break the tensor structure of the {\it l.h.s.}. 
Finally we have to calculate the Poisson bracket $\displaystyle{\{g(x)\mathop{,}^\otimes g\}}$. To this end we have recourse to the formula
\begin{eqnarray}
\{g(x)\mathop{,}^\otimes g(y)\} =
{\partial g(x)\over \partial {\cal G}^I(x)}\{{\cal G}^I(x)\mathop{,}^\otimes {\cal G}^J(y)\}{\partial g(y)\over \partial {\cal G}^J(y)}.   \label{defPoi}
\end{eqnarray}
By means of the Poisson brackets (\ref{Poisson}) it reads
\begin{eqnarray}
\{g(x)\mathop{,}^\otimes g(y)\} = {2\pi\over k}\Big[\theta(x-y)t_{AB}^+\delta^Ag(x)\otimes \delta^Bg(y)
-\theta(y-x)t_{AB}^+\delta^Ag(y)\otimes\delta^Bg(x)\Big].  \nonumber\\   
\label{Poi2} 
\end{eqnarray}
Plug this Poisson bracket into the {\it r.h.s.} of (\ref{gg}). First of all note that  (\ref{gg}) may be put into a simplified form 
\begin{eqnarray}
&\ & \{g(x)\mathop{,}^\otimes T_{++}(y)\}  \nonumber\\
&\ &\quad\quad 
= k\Bigg({\rm Tr}[\partial_y(\{g(x)\mathop{,}^\otimes g\}g^{-1})
(\partial_y gg^{-1})]+{2\over \sqrt 3}\partial_y^2{\rm Tr}[\{g(x)\mathop{,}^\otimes g\}g^{-1}T^Y]\Bigg), \quad\quad \label{gg'}
\end{eqnarray}
as follows. The quntities $\delta g$ and  $\delta gg^{-1}$ by the transformation (\ref{SL(3)})  have the matrix form 
\begin{eqnarray}
\left(
\begin{array}{ccc}
\hspace{0.1cm} *\hspace{0.15cm}  & 0\hspace{0.2cm} & 0\hspace{0.1cm} \\ 
\hspace{0.1cm} {*}\hspace{0.15cm}  & *\hspace{0.2cm} & *\hspace{0.1cm} \\
\hspace{0.1cm} {*}\hspace{0.15cm} & *\hspace{0.2cm}& * \hspace{0.1cm}
\end{array}\right).    \nonumber
\end{eqnarray}
So does the Poisson bracket $\displaystyle{\{g(x)\mathop{,}^\otimes g\}g^{-1}}$  which is calculated by  (\ref{Poi2}). 
Therefore we can  simplify  the second term of (\ref{gg}) 
as 
\begin{eqnarray}
\partial_y{\rm Tr}[\partial_y(\{g(x)\mathop{,}^\otimes g\}g^{-1})gT^Yg^{-1}]
&=& \partial_y{\rm Tr}[\partial_y(\{g(x)\mathop{,}^\otimes g\}g^{-1})g_LT^Yg_L^{-1}]   \nonumber\\
&=&\partial_y^2{\rm Tr}[\{g(x)\mathop{,}^\otimes g\}g^{-1}T^Y]  \nonumber
\end{eqnarray}
to find (\ref{gg'}). Next we remember  that $\delta^B T_{++}(y)=0$. Owing to this invariance  the {\it r.h.s.} of (\ref{gg'}) is vanishing except when the derivative $\partial_y$ acts on the step functions $\theta(x-y)$ and $\theta(y-x)$. Hence picking up both contributions we get 
\begin{eqnarray}
\{g(x) \mathop{,}^\otimes T_{++}(y)\} &=& 4\pi\Bigg(\partial_y\theta(x-y)t_{AB}\delta^Ag(x)\otimes {\rm Tr}[(\delta^Bgg^{-1}) (\partial_ygg^{-1})] \nonumber\\
&\ & \hspace{1cm}+{2\over \sqrt 3} \partial_y^2\theta(x-y)t_{AB}\delta^Ag(x)\otimes {\rm Tr}[(\delta^Bgg^{-1})T^Y]   \nonumber\\
&\ & \hspace{1.5cm}+{4\over \sqrt 3} \partial_y\theta(x-y)t_{AB}\delta^Ag(x)\otimes \partial_y{\rm Tr}[(\delta^Bgg^{-1})T^Y]\ \Bigg).\ \ \ \ \ \   \label{gT'}
\end{eqnarray}
Here note that  $t_{AB}^+$ could be  changed  to the usual Killing metric $t_{AB}$. 
We evaluate  the Poisson bracket (\ref{TT}) by plugging  this expression for $\displaystyle{\{g\mathop{,}^\otimes T_{++}(y)\}} $. The second term of (\ref{TT})
 may be simplified similarly to that of (\ref{gg}). 
Again due to the invariance $\delta^AT_{++}(x)=0$ there contribute only the terms with $\theta(x-y)$ differentiated by $x$. By using these facts we calculate the Poisson bracket (\ref{TT}) term  by term. Then the contribution from the  first term of (\ref{gT'}) reads 
\begin{eqnarray}
&\ & \{T_{++}(x) \mathop{,}^\otimes T_{++}(y)\}_{\rm first\ term} \nonumber\\
&\ & \hspace{1cm}=4\pi k\Bigg( \partial_x\partial_y\theta(x-y)t_{AB}
{\rm Tr}[(\partial_xgg^{-1})(\delta^Agg^{-1}) ]\otimes{\rm Tr}[(\delta^Bgg^{-1}) (\partial_ygg^{-1})] \nonumber\\
&\ & \hspace{2cm}+{2\over\sqrt 3} \partial_x^2\partial_y\theta(x-y)t_{AB}{\rm Tr}[(\delta^Agg^{-1})
T^Y]\otimes {\rm Tr}[(\delta^Bgg^{-1})(\partial_ygg^{-1})]  \nonumber\\
&\ & \hspace{2cm}+{4\over\sqrt 3} \partial_x\partial_y\theta(x-y)t_{AB}\partial_x{\rm Tr}[(\delta^Agg^{-1})T^Y]\otimes{\rm Tr}[(\delta^Bgg^{-1})(\partial_ygg^{-1})]\Bigg).\quad\ \ \  \label{TT1'}
\end{eqnarray}
The contribution from the second term of (\ref{gT'}) reads
\begin{eqnarray}
&\ & \{T_{++}(x) \mathop{,}^\otimes T_{++}(y)\}_{\rm second\ term} \nonumber\\
&\ & \hspace{1cm}=4\pi k{2\over\sqrt 3}\Bigg( \partial_x\partial_y^2\theta(x-y)t_{AB}
{\rm Tr}[(\partial_xgg^{-1})(\delta^Agg^{-1}) ]\otimes{\rm Tr}[(\delta^Bgg^{-1}) T^Y] \nonumber\\
&\ & \hspace{2cm}+{2\over\sqrt 3} \partial_x^2\partial_y^2\theta(x-y)t_{AB}{\rm Tr}[(\delta^Agg^{-1})
T^Y]\otimes {\rm Tr}[(\delta^Bgg^{-1})T^Y]  \nonumber\\
&\ & \hspace{2cm}+{4\over\sqrt 3} \partial_x\partial_y^2\theta(x-y)t_{AB}\partial_x{\rm Tr}[(\delta^Agg^{-1})T^Y]\otimes{\rm Tr}[(\delta^Bgg^{-1})T^Y]\Bigg).\quad   \label{TT2}
\end{eqnarray}
But there is no contribution from the third term as can be seen as follows. 
 Namely note that 
\begin{eqnarray}
\delta gg^{-1}=\epsilon\cdot T-gug^{-1}=\epsilon\cdot T-(\epsilon_L^1g_LT_{R1}g_L^{-1}+\epsilon_L^2g_LT_{R2}g_L^{-1}),
  \label{gzero}
\end{eqnarray}
by writing the transformation (\ref{SL(3)}) in the infinitesimal form. From this we know that $\partial_y {\rm Tr}[(\delta^Bgg^{-1}) T^Y]$ in the last term of (\ref{gT'})   makes no  contraction with ${\rm Tr}[(\delta^Agg^{-1}) T^Y]$ and ${\rm Tr}[(\delta^Agg^{-1})(\partial_xgg^{-1}) ]$ because the latter quantities do not contain components along the variations $\epsilon_R^1,\epsilon_R^2$. So  the last term of (\ref{gT'}) does not contribute. For the same reason the last terms of  (\ref{TT1'}) and (\ref{TT2}) do not contribute either, and the remaining terms  may be calculated by simply  setting $\delta^Bgg^{-1}$ to be $T^B$. As the result we find 
 the Virasoro algebra
\begin{eqnarray}
{1\over 2\pi}\int dx\eta(x) \{T_{++}(x) \mathop{,}^\otimes T_{++}(y)\}=
\eta(y)\partial_yT_{++}(y)+2\Big(\partial_y\eta(y)\Big)T_{++}(y) -{2k\over 3} \partial_y^3\eta(y).  \nonumber \\
 \label{Virasoroo}
\end{eqnarray}

\subsection{A classical exchange algebra}

For the constrained $SL(3)$ WZWN model there also exists a quantity such as (\ref{primary2}), called  $SL(3)$ conformal primary. It takes the form  
\begin{eqnarray}
\Psi= {1\over \Delta}
\left(
\begin{array}{c}
1 \\
G_1  \\ 
G_2  
\end{array}
\right).
 \label{primary3}
\end{eqnarray}
 It indeed
linearly transforms   by 
    (\ref{G-trans}) and (\ref{lambda}) as 
\begin{eqnarray}
\delta\Psi=\epsilon\cdot T\Psi,   \label{linear2}
\end{eqnarray}
 with the  generators (\ref{gene}) in the fundamental representation of $SL(3)$. This can be shown   by writing (\ref{gzero}) as
\begin{eqnarray}
\delta (g_Lg_0) &=&\epsilon\cdot T(g_Lg_0) - (g_Lg_0)u.  \nonumber  \end{eqnarray}
We calculate both sides by using the explicit form (\ref{para1}) and (\ref{trans2}). Then the first column vector of this matrix equation gives the transformation (\ref{linear2}). Using linearity of this transformation as well as 
the formula (\ref{defPoi}) with $g$ replaced by $\Psi$ we can show  a classical exchange algebra for the $SL(3)$ conformal primary
\begin{eqnarray}
\{\Psi(x)\mathop{,}^\otimes \Psi(y)\}={2\pi\over k}[\theta(x-y)r^+ + \theta(y-x) r^-]\Psi(x)\otimes\Psi(y),   \label{CEA}
\end{eqnarray}
in which  use  is made of  $t_{AB}^+=-t_{BA}^-$.

Moreover  we can show that the $SL(3)$ conformal primary (\ref{primary3}) transforms with the weight $-{2\over 3}$ in the right-moving sector as
\begin{eqnarray}
{1\over 2\pi}\int dx\ \eta(x)\{T_{++}(x)\mathop{,}^\otimes \Psi(y)\}=\eta(y)\partial_y\Psi(y) -{2\over 3}(\partial_y\eta(y))\Psi(y). \label{Psi}
\end{eqnarray}
To  this end it is necessary to examine the transformation property of ${\cal G}^I=(G_1,G_2, F,G,\lambda,$ $\mu)$, which $\Psi$ is composed of. We use (\ref{gT'}) again. Note that it holds  even if  $g$ is replaced by ${\cal G}^I$. 
To evaluate the {\it r.h.s.} of (\ref{gT'}) note also that we have the formulae
\begin{eqnarray}
{\rm Tr}[(\delta^Bgg^{-1})(\partial_ygg^{-1})]&=&{\rm Tr}[(\delta^Bg_0g_0^{-1})(\partial_yg_0g_0^{-1})],   \nonumber\\
{\rm Tr}[(\delta^Bgg^{-1})T^Y]&=&{\rm Tr}[(\delta^Bg_0g_0^{-1})T^Y],   \nonumber
\end{eqnarray}
by putting $g=g_Lg_0$.  
Calculate the {\it r.h.s.} of these formulae using the infinitesimal form of (\ref{2nd}) we find 
\begin{eqnarray}
&\ &t_{AB}\epsilon^A{\rm Tr}[(\delta^Bgg^{-1})(\partial_ygg^{-1})]=
  {\rm Tr}[\rho\cdot \hat H (\partial_yg_0g_0^{-1})]    \nonumber\\
&\ &\hspace{3cm} ={\epsilon^Q\over 2}{1\over \Delta}(\mu \partial_y\lambda-G\partial_yF+F\partial_yG-\lambda \partial_y\mu)
-{\sqrt 3\over 2} \epsilon^Y{\partial_y\Delta\over \Delta}
 \nonumber\\
&\ &\hspace{3cm}-{\epsilon_L^1\over \sqrt 2}{1\over \Delta}\Big\{G_1(\mu \partial_y\lambda-G\partial_y F+ {\partial_y\Delta\over \Delta})+G_2(-F\partial_y\lambda+\partial_y F) \Big\} \nonumber\\
&\ &\hspace{3cm}-{\epsilon_L^2\over \sqrt 2}{1\over \Delta}\Big\{G_1(\mu \partial_yG-G\partial_y \mu)+G_2(-F\partial_yG+\lambda\partial_y\mu+{\partial_y\Delta\over \Delta}) \Big\} \nonumber\\
&\ &\hspace{3cm}+{\epsilon_L^3\over \sqrt 2}{1\over \Delta}(\mu\partial_yG-G\partial_y\mu)+{\epsilon_{R3}\over \sqrt 2}{1\over \Delta}(-F\partial_y\lambda+\lambda \partial_yF),     \label{rela2} \\
&\ &\hspace{1cm}\ t_{AB}\epsilon^A{\rm Tr}[(\delta^Bgg^{-1})T^Y]=
{\rm Tr}[\rho\cdot \hat H T^Y]    \nonumber \\
&\ &\hspace{3cm}={\epsilon^Y\over  2}+{1\over 2}\sqrt{3\over 2}(\epsilon_L^1G_1+\epsilon_L^2G_2).  \label{rela1} 
\end{eqnarray}
Plugging them into (\ref{gT'}) and making contraction with respect the indices $A$ and $B$ according to the definition of the Casimir (\ref{norm}) yields 
\begin{eqnarray}
{1\over 2\pi}\int dx\ \eta(x)\{T_{++}(x)\mathop{,}^\otimes {\cal G}^I(y)\}=\eta(y)\partial_y{\cal G}^I(y) +h_{{\cal G}^I}(\partial_y\eta(y)){\cal G}^I(y),  \label{conf'}
\end{eqnarray}
in which 
\begin{eqnarray}
h_{G_1} = h_{G_2} = 0,  \quad\quad 
h_\lambda =h_F=h_G=h_\mu={1\over 3}.   \nonumber
\end{eqnarray}
These conformal weights sum up to give $-{2\over 3}$ to $\Psi$. 
A by-product of this result is  that the constrained currents $J_{+1}$ and $J_{+2}$, defined by (\ref{current}), no longer have   weight 1, but 0 in the right-moving sector. Owing to this twisting of the conformal weight it is justified {\it a posteriori}   to  constrain $J_{+1}$ and $J_{+2}$ to be constant  as (\ref{current}).

\vspace{1cm}

\section{Irreducibly constrained $SL(N)$ WZWN models with $N\ge 4$}
\setcounter{equation}{0}

\subsection{Generalization from $SL(3)$}

The arguments so far given can be straightforwardly extended to more general cases. Namely 
the $SL(M+N)$ WZWN model may be gauged by choosing the gauge field in a general form as
\begin{eqnarray}
A_-\equiv \left(
\begin{array}{c|c}
 a_{-\alpha}^{\ \ \beta}  &  a_{-\alpha}^{\ \ j} \\
\hline
 a_{-i}^{\ \ \beta}  & a_{-i}^{\ \ j} 
\end{array}\right)
=\left(
\begin{array}{c|c}
\begin{array}{ccc}
0 & \cdots & 0  \\
\vdots & \ddots & \vdots \\
0 & \cdots & 0  \\
\noalign{\vskip0.1cm}
\end{array}   &  
\begin{array}{ccc}
a_{-1}^{\ \ 1} & \cdots\cdots & a_{-1}^{\ \ N}  \\
\vdots & \ddots\ddots & \vdots \\
a_{-M}^{\ \ 1} & \cdots\cdots & a_{-M}^{\ \ N}  \\
\noalign{\vskip0.1cm}
\end{array}  \\
\hline
\begin{array}{ccc}
\noalign{\vskip0.1cm}
0 & \cdots & 0  \\
\vdots & \ddots & \vdots \\
\vdots & \ddots & \vdots \\
0 & \cdots & 0  
\end{array}   &
\begin{array}{ccc}
\noalign{\vskip0.1cm} 
0 \ \  & \cdots\cdots &\ \  0  \\
\vdots\ \  & \ddots\ddots &\ \  \vdots \\
\vdots\ \  & \ddots\ddots &\ \  \vdots \\
0\ \  & \cdots\cdots &\ \  0 
\end{array}  
\end{array}\right).    \label{block}
\end{eqnarray}
The equation of motion for $A_-$ yields the constraints
\begin{eqnarray}
J_{+k}^{\ \ \ \gamma}={\rm Tr}[g^{-1}\partial_+g T_{Rk}^{\ \ \ \gamma}]={\rm const.},
 \label{c-currents}
\end{eqnarray}
in which  $T_{Rk}^{\ \ \ \gamma}$ are generators represented by  a  block matrix form (\ref{block})  as
\begin{eqnarray}
(T_{Rk}^{\ \ \ \gamma})_\alpha^{\  j}\propto\left(
\begin{array}{c|c}
\ \  0 \ \  &   \delta^j_k\delta^\gamma_\alpha \\
\hline
 \ \ 0 \ \  &   0
\end{array}\right),\quad  j,k=1,\cdots,N,\quad  \alpha,\gamma=1,\cdots,M.
\end{eqnarray}
The gauge-fixed group element $g=g_Lg_0$ is parametrized as
\begin{eqnarray}
g_L&=&\left(
\begin{array}{c|c}
 1  & \ \ 0\ \ \\
\hline
 G_i^{\  \beta}  & \ \  1\ \
\end{array}\right),    \\
g_0&=&(\det\lambda'\lambda)^{-{1\over M+N}}\left(
\begin{array}{c|c}
 {\lambda'}_\alpha^{\ \beta}  & \ \ 0\ \ \\
\hline
 0  &  \lambda_i^{\ j} 
\end{array}\right)
\equiv
\left(
\begin{array}{c|c}
 {\Lambda'}_\alpha^{\ \beta}  & \ \ 0\ \ \\
\hline
 0  &  \Lambda_i^{\ j}   
\end{array}\right).        \label{defL}
\end{eqnarray}
Let  $G_i^{\  \beta}, {\Lambda'}_\alpha^{\ \beta}, \Lambda_i^{\ j}$  in  the parametrization denote by ${\cal G}^I$ again. They are transformed  by the relation (\ref{SL(3)}), i.e.,
\begin{eqnarray}
g({\cal G})\longrightarrow e^{\epsilon\cdot T}g({\cal G}) U_R^{-1}=g({\cal G}'),  \label{global3} 
\end{eqnarray}
with $g({\cal G})=g_L(G)g_0(\Lambda',\Lambda)$ and 
\begin{eqnarray}
e^{\epsilon\cdot T}\in SL(M+N),\quad\quad\quad
U_R^{-1}=\left(
\begin{array}{c|c}
 1 & * \\
\hline
 0 & 1
\end{array}\right)\equiv 1 -u.      \nonumber
\end{eqnarray}
It is shown exactly in the same way as for the case of $SL(3)$ that 
the constrained currents (\ref{c-currents}) and the $SL(M+N)$ WZWN action 
are invariant by the transformation (\ref{global3}). 
 The symmetry is realized by the coset space  $SL(M+N)/\{SL(M)\otimes SL(N)\otimes U(1)\}$, which is irreducible again. 
We find a traceless $U(1)$ generator by solving the equation $[u, T^{U(1)}]= -u$ similarly to the case of  $SL(3)$. It is given by 
\begin{eqnarray}
T^{U(1)}= {N\over M+N}
\left(
\begin{array}{c|c}
1 & 0 \\
\hline 
0 & 0 
\end{array}\right) - {M\over M+N}
\left(
\begin{array}{c|c}
0 & 0 \\
\hline 
0 & 1 
\end{array}\right).  \label{TU}
\end{eqnarray}
With  this  $U(1)$ generator at hand the whole arguments of sections 3 go through. To be concrete, the normalization of the $SL(M+N)$ Lie algebra 
 is done by requiring the quadratic Casimir 
$$
-C\equiv t_{AB}T^AT^B=T_{Ri}^{\ \ \beta}T_{L\beta}^{\ \  i}+T_{L\beta}^{\ \ i}T_{Ri}^{\ \ \beta}
+ {T'}_\alpha^{\  \beta}{T'}_\beta^{\  \alpha} + T_i^{\  j}T_j^{\  i}
+T^YT^Y,
$$
to be $-(M+N)$ in the adjoint representation. Then $T_L(T_R)$ takes in a generalized form  from (\ref{T}) and (\ref{gene}) such as
\begin{eqnarray}
[T_{Ri}^{\ \ \beta}]_\gamma^{\ k}={1\over \sqrt 2}\delta_i^k\delta_\gamma^\beta,\quad\quad 
[T_{L\beta}^{\ \  i}]_k^{\ \gamma}={1\over \sqrt 2}\delta^i_k\delta^\gamma_\beta,      
\end{eqnarray}
while  the $SL(M)(SL(N))$ generators $T'(T)$ and $T^Y$  are given by 
\begin{eqnarray}
[{T'}_\alpha^{\  \beta}]_\gamma^{\ \lambda}&=&{1\over \sqrt 2}(\delta_\alpha^\lambda\delta_\gamma^\beta-{1\over M}\delta_\alpha^\beta\delta_\gamma^\lambda), \quad\quad [T_i^{\  j}]_k^{\ l}={1\over \sqrt 2}(\delta_i^l\delta_k^j-{1\over N}\delta_i^j\delta_k^l),  \nonumber\\
   T^Y &=& {\sqrt {M+N\over 2MN}}T^{U(1)}.
\end{eqnarray}
The concrete forms of the Killing vectors $\delta^A{\cal G}^I$ 
 are obtained from the transformation (\ref{global3}) by similar calculations  to the case of $SL(3)$. Parameterizing  $e^{\epsilon\cdot T}$  as
\begin{eqnarray}
\epsilon\cdot T &=&\epsilon_{Ri}^{\ \ \beta}T_{L\beta}^{\ \  i}+\epsilon_{L\beta}^{\ \ i}T_{Ri}^{\ \ \beta}
+ {\epsilon'}_\alpha^{\  \beta}{T'}_\beta^{\  \alpha} + \epsilon_i^{\  j}T_j^{\  i}
+\epsilon^YT^Y    \nonumber\\
&=& {1\over \sqrt 2}\left(
\begin{array}{c|c}
\ \epsilon' \ & \  \epsilon_L\  \\
\hline
 \ \epsilon_R \
    & \   \epsilon\ \
\end{array}\right) + \epsilon^YT^Y,
\end{eqnarray}
in the block matrix form (\ref{block})
we find the compensators to be 
\begin{eqnarray}
\rho\cdot \hat H &=& 
{1\over \sqrt 2}\left(
\begin{array}{c|c}
\begin{array}{c}
\vspace{0.2cm}
 [\epsilon'+\epsilon_L G]_\alpha^{\ \beta}
\end{array} 
& 
\ \ \begin{array}{c}
\vspace{0.2cm}
[\epsilon_L]_\alpha^{\ j}
\end{array} 
 \ \ \\
\hline
\begin{array}{c}
\noalign{\vskip0.2cm}
 0
\end{array} 
  & \ \ 
\begin{array}{c}
\noalign{\vskip0.2cm}
 [\epsilon-G\epsilon_L]_i^{\ j }
\end{array}
 \ \
\end{array}\right) +\epsilon^YT^Y, \label{rho}\\
\noalign{\vskip0.2cm}
u &=& 
{1\over \sqrt 2}\left(
\begin{array}{c|c}
\begin{array}{c}
\vspace{0.2cm}
\hspace{0.95cm} 0\hspace{0.9cm}
\end{array} 
& 
\ \ \begin{array}{c}
\vspace{0.2cm}
[{\Lambda'}^{-1}\epsilon_L\Lambda]_\alpha^{\ j}
\end{array} 
 \ \ \\
\hline
\begin{array}{c}
\noalign{\vskip0.2cm}
 0
\end{array} 
  & \ \ 
\begin{array}{c}
\noalign{\vskip0.2cm}
 0
\end{array}
 \ \
\end{array}\right).  
  \label{U}
\end{eqnarray}
The Killing vectors $\delta^A{\cal G}^I$ are given by 
\begin{eqnarray}
\delta G_i^{\ \beta}&=&{1\over \sqrt 2}\Bigg[\epsilon_R+\epsilon G-G\Bigg(\sqrt{M+N\over MN}\epsilon^Y+\epsilon'+\epsilon_L G\Bigg)\Bigg]_i^{\ \beta},  \nonumber\\
\delta{\Lambda'}_\alpha^{\ \beta}&=&{1\over \sqrt 2}\Bigg[\Bigg({N\over \sqrt{MN(M+N)}}\epsilon^Y+\epsilon'+\epsilon_LG\Bigg)\Lambda'\Bigg]_\alpha^{\ \beta},  \label{Killing3}\\
\delta\Lambda_i^{\ j}&=&{1\over \sqrt 2}\Bigg[\Bigg(-{M\over \sqrt{MN(M+N)}}\epsilon^Y+\epsilon-G\epsilon_L\Bigg)\Lambda\Bigg]_i^{\ j}.  \nonumber
\end{eqnarray}
The modified energy-momentum tensor, which is invariant by the transformation (\ref{global3}), takes the same form as (\ref{modT}) with $T^{U(1)}$ replaced by (\ref{TU}),  i.e.,
\begin{eqnarray}
T_{++}&=& k\Bigg({1\over 2}{\rm Tr}(g^{-1}\partial_+g)^2+\sqrt{2MN\over M+N}\partial_+{\rm Tr}[(g^{-1}\partial_+g)T^Y]\Bigg).   \nonumber
\end{eqnarray}

We set up the Poisson brackets and examine the Virasoro algebra for this energy-momentum tensor. The whole arguments for the case of $SL(3)$, given in subsection 3.2,  can be generalized straightforwardly. We are led  to  
 find the Virasoro algebra 
\begin{eqnarray}
{1\over 2\pi}\int dx\eta(x) \{T_{++}(x) \mathop{,}^\otimes T_{++}(y)\}=
\eta(y)\partial_yT_{++}(y)+2\Big(\partial_y\eta(y)\Big)T_{++}(y) -{MN\over M+N}k \partial_y^3\eta(y).\ \    \nonumber
\end{eqnarray}
The $SL(M+N)$ conformal primary $\Psi$ which transforms as $\delta\Psi=\epsilon\cdot T \Psi$ is also   found to be  
\begin{eqnarray}
\Psi= (\det\Lambda'\Lambda)^{-{1\over M+N}}
\left(
\begin{array}{c}
\begin{array}{c}
  [\Lambda' ]_\alpha^{\ \beta}
\end{array}   \\ 
\begin{array}{c}
\noalign{\vskip0.2cm}
 [ G\Lambda']_i^{\ \beta}  
\end{array} 
\end{array}\right).
 \nonumber
\end{eqnarray}
We may take any of column vectors in this rectangular matrix as the $SL(M+N)$ conformal primary. 
Similarly to the case of $SL(3)$ it satisfies the classical exchange algebra (\ref{CEA}).

\subsection{Conformal weight of ${\cal G}^I$ and $J_{+ i}^{\ \ \beta}$}

We have already seen that the constrained currents in (\ref{c-currents}) are invariant by the $SL(M+N)$ transformation (\ref{global3}), and yet have not examined their conformal transformations. For imposing the constraints (\ref{c-currents}) consistently it is also crucially important that the constrained currents have conformal weight $0$. We shall give a proof for this fact after that for the case of $SL(3)$. First of all write the constrained currents as
$$
J_{+i}^{\ \ \beta}= {\rm Tr}[(g_L^{-1}\partial_xg_L)(g_0 T_{Ri}^{\ \ \beta}g_0^{-1})].
$$
Suppose that the group variables ${\cal G}^I(=G_i^{\  \beta}, {\Lambda'}_\alpha^{\ \beta}, \Lambda_i^{\ j})$ obey the conformal transformation 
\begin{eqnarray}
{1\over 2\pi}\int dx\ \eta(x)\{T_{++}(x)\mathop{,}^\otimes {\cal G}^I(y)\}=\eta(y)\partial_y{\cal G}^I(y) +h_{{\cal G}^I}(\partial_y\eta(y)){\cal G}^I(y),  \label{conf3}
\end{eqnarray}
in which 
\begin{eqnarray}
h_G=0,  \quad\quad 
h_{\Lambda'} =-{N\over M+N}, \quad\quad h_{\Lambda} ={M\over M+N}.
   \label{weight}
\end{eqnarray}
Then obviously $g_L^{-1}\partial_xg_L$ has  weight $1$, while $g_0 T_{Ri}^{\ \ \beta}g_0^{-1}$ reads 
\begin{eqnarray}
\left(
\begin{array}{c|c}
\begin{array}{c}
\vspace{0.2cm}
\hspace{0.95cm} 0\hspace{0.9cm}
\end{array} 
& \quad {\Lambda'}_i^{\ j}T_{Rj}^{\ \ \alpha} \Lambda^{-1\beta}_{\ \ \alpha} 
\ \ \begin{array}{c}
\vspace{0.2cm}

\end{array} 
 \ \ \\
\hline
\begin{array}{c}
\noalign{\vskip0.2cm}
 0
\end{array} 
  & \ \ 
\begin{array}{c}
\noalign{\vskip0.2cm}
 0
\end{array}
 \ \
\end{array}\right)
\end{eqnarray}
and has weight $h_{\Lambda'}-h_{\Lambda}=-1$ due to (\ref{weight}). Hence the weight of the constrained currents is $0$. Now it suffices to show (\ref{conf3}) with (\ref{weight}) to have this conclusion. As has been done to show (\ref{conf'}) 
we use  the formula (\ref{gT'}), which now reads
\begin{eqnarray}
\{{\cal G}^I(x) \mathop{,}^\otimes T_{++}(y)\} &=& 4\pi\Bigg(\partial_y\theta(x-y)t_{AB}\delta^A{\cal G}^I(x)
\otimes {\rm Tr}[(\delta^Bgg^{-1}) (\partial_ygg^{-1})] \nonumber\\
&\ & \hspace{0.6cm}+\sqrt{2MN\over M+N} \partial_y^2\theta(x-y)t_{AB}\delta^A{\cal G}^I(x)\otimes {\rm Tr}[(\delta^Bgg^{-1})T^Y]   \nonumber\\
&\ & \hspace{0.5cm}+2\sqrt{2MN\over M+N}  \partial_y\theta(x-y)t_{AB}\delta^A{\cal G}^I(x)\otimes \partial_y{\rm Tr}[(\delta^Bgg^{-1})T^Y]\ \Bigg).\ \ \ \ \ \ \ \ \label{gT}
\end{eqnarray}
We calculate the {\it r.h.s.} term by term.  The first term was the most hard part to calculate for the case of $SL(3)$. Here we give a rather simple calculation  for the general case. Indeed for ${\cal G}^I=G_i^{\ \beta}$ we can show that
$$
t_{AB}\delta^A G_i^{\ \beta}\otimes \delta^B {\cal G}^J=0\quad {\rm for\ all}\ \ J,
$$
by an explicit calculation by means of the Killing vectors (\ref{Killing3}).
 Therefore the first term with ${\cal G}^I=G_i^{\ \beta}$ is vanishing. For ${\cal G}^I={\Lambda'}_\alpha^{\ \beta}$ we calculate as 
\begin{eqnarray}
t_{AB}\delta^A{\Lambda'}_\alpha^{\ \beta}\otimes {\rm Tr}[(\delta^Bgg^{-1})
(\partial_ygg^{-1}]&=& t_{AB}\delta^A{\Lambda'}_\alpha^{\ \beta}\otimes {\rm Tr}[(\delta^Bg_0g_0^{-1})
(\partial_yg_0g_0^{-1})]     \nonumber\\
&=& {1\over 2}\otimes \Bigg[
\left(
\begin{array}{c|c}
 \Lambda'  & \ 0 \ \\
\hline
  0   & \  0\
\end{array}\right)
(\partial_yg_0g_0^{-1})\Bigg]_\alpha^{\ \beta}={1\over 2}\otimes\partial_y{\Lambda'}_\alpha^{\ \beta},   \nonumber
\end{eqnarray}
by finding $\delta^Bg_0g_0^{-1}$ from $\rho\cdot \hat H$ as (\ref{rho})
 and making it  contract with $\delta^A {\Lambda'}_\alpha^{\ \beta}$. Here $\epsilon^Y$-component of  $\rho\cdot \hat H$ does not contribute due to the traceless condition
 $$
{\rm Tr}[\partial_y{\Lambda'}{\Lambda'}^{-1}+\partial_y{\Lambda}{\Lambda}^{-1} ]=0,
$$
which follows from the definition (\ref{defL}), i.e., $\det{\Lambda'\Lambda}=1$. The calculation for the case of ${\cal G}^I=\Lambda_i^{\ j}$ can be  similarly done. 
Altogether the first term of (\ref{gT}) reduces to  
\begin{eqnarray}
&\ &\partial_y\theta(x-y)t_{AB}\delta^A{\cal G}^I(x)    
\otimes {\rm Tr}[(\delta^Bgg^{-1}) (\partial_ygg^{-1})]  \nonumber\\
&\ & \hspace{2cm} =\left\{
\begin{array}{rl}
0, & \quad\mbox{for ${\cal G}^I(x)=G_i^{\   \beta}(x)$}, \\
 -\delta(x-y){1\over 2}\otimes\partial_y {\Lambda'}_\alpha^{\   \beta}(y), &
   \quad\mbox{for ${\cal G}^I(x)={\Lambda'}_\alpha^{\   \beta}(x)$ },\\
-\delta(x-y){1\over 2}\otimes\partial_y {\Lambda}_i^{\   j}(y),  & 
   \quad\mbox{for ${\cal G}^I(x)={\Lambda}_i^{\   j}(x)$}.
\end{array}\right.
\end{eqnarray}
As for the second and third terms in  (\ref{gT}) we calculate them as 
\begin{eqnarray}
t_{AB}\delta^AG_i^{\ \beta}(x)\otimes {\rm Tr}[(\delta^Bgg^{-1})T^Y] &=&  
-{1\over 2}\sqrt{{M+N}\over 2MN}[G(x)\otimes 1-1\otimes G(y)]_i^{\ \beta}, \nonumber\\
t_{AB}\delta^A{\Lambda'}_\alpha^{\ \beta}(x)\otimes {\rm Tr}[(\delta^Bgg^{-1})T^Y] 
&=& {N\over 2\sqrt{2MN(M+N)}}{\Lambda'}_\alpha^{\ \beta}(x)\otimes 1, 
 \nonumber\\
t_{AB}\delta^A\Lambda_i^{\ j}(x)\otimes {\rm Tr}[(\delta^Bgg^{-1})T^Y] &=&      -{M\over 2\sqrt{2MN(M+N)}}\Lambda_i^{\ j}(x)\otimes 1, 
 \nonumber
\end{eqnarray}
by the formula 
\begin{eqnarray}
t_{AB}\epsilon^A{\rm Tr}[(\delta^Bgg^{-1})T^Y]={\rm Tr}[\rho\cdot\hat H T^Y]={\epsilon^Y\over 2}+{1\over 2}\sqrt{M+N\over MN}{\rm Tr}[\epsilon_LG],    \label{deltaB}
\end{eqnarray}
which follows from (\ref{rho}) directly. Putting these results into (\ref{gT}) together 
we can verify the conformal transformation  (\ref{conf3}) with (\ref{weight}).

As the result the constrained currents $J_{+i}^{\ \beta}$  have weight $0$. Therefore its conformal transformation reads
\begin{eqnarray}
{1\over 2\pi}\int dx\ \eta(x)\{T_{++}(x)\mathop{,}^\otimes J_{+i}^{\ \ \beta}(y) \}=\eta(y)\partial_yJ_{+i}^{\ \ \beta}(y),    \label{J+}
\end{eqnarray}
which is vanishing upon imposed the constraints (\ref{c-currents}). It is worth  demonstrating this equation more directly. Using (\ref{variation}) we calculate  the 
Poisson bracket as
\begin{eqnarray}
\{J_{+i}^{\ \ \beta}(x)\mathop{,}^\otimes T_{++}(y)\} =  
{\rm Tr}[\partial_x(\{g(x)\mathop{,}^\otimes T_{++}(y)\}g^{-1})gT_{Ri}^{\ \ \beta}g^{-1}].    \label{JT}
\end{eqnarray}
Plug (\ref{gT'}) into the {\it r.h.s.}, after generalizing it for the case of $SL(M+N)$ as (\ref{gT}). Keep only the terms with $\theta(x-y)$ differentiated by $x$ because other terms drop out due to the invariance $\delta^AJ_{+i}^{\ \ \beta}(x)=0$. Then (\ref{JT}) becomes  
\begin{eqnarray}
&\ &\{J_{+i}^{\ \ \beta}(x)\mathop{,}^\otimes T_{++}(y)\} \nonumber  \\
&\ &\hspace{0.5cm} = 4\pi\Bigg(\partial_x\partial_y\theta(x-y)t_{AB}{\rm Tr}[(\delta^Agg^{-1})
gT_{Ri}^{\ \ \beta}g^{-1}]\otimes {\rm Tr}[(\delta^Bgg^{-1}) (\partial_ygg^{-1})] \nonumber\\
&\ & \hspace{2cm}+\sqrt{2MN\over M+N}\partial_x\partial_y^2\theta(x-y)t_{AB}
{\rm Tr}[(\delta^Agg^{-1})
gT_{Ri}^{\ \ \beta}g^{-1}]
\otimes {\rm Tr}[(\delta^Bgg^{-1})T^Y]   \label{JT'}\\
&\ & \hspace{2cm}+2\sqrt{2MN\over M+N} \partial_x\partial_y\theta(x-y)t_{AB}
{\rm Tr}[(\delta^Agg^{-1})
gT_{Ri}^{\ \ \beta}g^{-1}]
\otimes \partial_y{\rm Tr}[(\delta^Bgg^{-1})T^Y]\ \Bigg).\ \ \ \   \nonumber
\end{eqnarray}
Writing again the transformation (\ref{global3}) in the infinitesimal form 
$$
\delta gg^{-1} =\epsilon\cdot T -gug^{-1}
$$ 
and using (\ref{U}),  we have
\begin{eqnarray}
t_{AB}\epsilon^B{\rm Tr}[(\delta^Agg^{-1})gT_{Ri}^{\ \ \beta}g^{-1}]&=& 
 {\rm Tr}[\epsilon\cdot TgT_{Ri}^{\ \ \beta}g^{-1}],  \nonumber \\
t_{AB}\epsilon^A{\rm Tr}[(\delta^Bgg^{-1}) (\partial_ygg^{-1})]
 &=& {\rm Tr}[\epsilon\cdot T(\partial_ygg^{-1})] -{1\over \sqrt 2}{\rm Tr}
[\left(
\begin{array}{c|c}
\ 0 \ & \  \epsilon_L\  \\
\hline
 \ 0 \
    & \ 0 \ \
\end{array}\right) 
g_L^{-1}\partial_yg_L]     \nonumber\\
&=& {\rm Tr}[\epsilon\cdot T(\partial_ygg^{-1})]-{1\over \sqrt 2}\partial_y{\rm Tr}[\epsilon G]. 
\label{finalcal}   
\end{eqnarray}
Making contraction of these two quantities reduces the first term of (\ref{JT'}) to 
\begin{eqnarray}
&\ & \partial_x\partial_y\theta(x-y)t_{AB}{\rm Tr}[(\delta^Agg^{-1})
gT_{Ri}^{\ \ \beta}g^{-1}]\otimes {\rm Tr}[(\delta^Bgg^{-1}) (\partial_ygg^{-1})]   \nonumber\\
&\ &\hspace{1cm}= \partial_x\partial_y\theta(x-y)\Bigg({1\over 2}{\rm Tr}[gT_{Ri}^{\ \ \beta}g^{-1}\otimes \partial_ygg^{-1}]    \nonumber\\
&\ & \hspace{4cm}
-\sqrt{2MN\over M+N}t_{AB}{\rm Tr}[(\delta^Agg^{-1})
gT_{Ri}^{\ \ \beta}g^{-1}]\otimes \partial_y{\rm Tr}[(\delta^Bgg^{-1})T^Y] \Bigg),
\nonumber
\end{eqnarray}
by  our normalization ${\rm Tr} T^AT^B={1\over 2}t^{AB}$ and (\ref{deltaB}). Using (\ref{JT'}), of which first term is replaced by this equation, we perform the integration of the {\it l.h.s.} of (\ref{J+}) to find the {\it r.h.s.}.

\vspace{1cm}

\renewcommand{\theequation}{\thesection.\arabic{equation}}

\section{Reducibly constrained $SL(N)$ WZWN models}
\setcounter{equation}{0}

So far we have discussed assuming that the gauge-fixed symmetry was irreducible. Finally we extend the arguments to reducible cases. Then the relevant coset space of the symmetry is  $SL(N)/\{S\otimes U(1)^l\}(\subseteq SL(N)/ U(1)^{N-1})$ with some subgroup $S$. In this section we discuss the  largest case, i.e., $SL(N)/ U(1)^{N-1}$. For that case we choose the gauge field as 
\begin{eqnarray}
A_-=\left(
\begin{array}{ccccc}
\hspace{0.1cm} 0      &\hspace{0.1cm} a_{-1}^{\ \ 2} &\hspace{-0.1cm} a_{-1}^{\ \ 3} & \cdots   & a_{-1}^{\ \ N}  \\
\hspace{0.1cm} 0      &\hspace{0.1cm}   0       &\hspace{-0.1cm} a_{-2}^{\ \ 3} & \cdots   & a_{-2}^{\ \ N}  \\
\hspace{0.1cm} 0      &\hspace{0.1cm}   0       &\hspace{-0.1cm}   0     
  & \cdots   & a_{-3}^{\ \ N}  \\
\hspace{0.1cm} \vdots &\hspace{0.1cm}\vdots    &\hspace{-0.1cm} \vdots    
& \ddots    & \vdots     \\
\hspace{0.1cm} 0      &\hspace{0.1cm}   0       &\hspace{-0.1cm}   0    & \cdots   & 0         \\
\end{array}\right).
\end{eqnarray}
The equation of motion for $A_-$ gives the currents 
\begin{eqnarray}
J_{+i}^{\ \  j}={\rm Tr}[g^{-1}\partial_+g T_{Ri}^{\ \  j}], \quad\quad 
 j< i=2,3,\cdots,N,   \label{redcurr}
\end{eqnarray}
with $(T_{Ri}^{\ \  j})_k^{\ \ l}=\delta_i^l\delta_k^j/\sqrt 2$.  We fix the gauge as
\begin{eqnarray}
g_L&=& 
\left(
\begin{array}{ccccc}
 1   \hspace{-0.1cm}   & 0\hspace{0.1cm} & 0 &\hspace{0.05cm} \cdots   & 0 \ \\
 G_2^{\ 1} \hspace{-0.1cm}     &   1\hspace{0.05cm}       & 0 &\hspace{0.05cm} \cdots   & 0  \hspace{0.1cm} \\
 G_3^{\ 1}  \hspace{-0.1cm}    &    G_3^{\ 2}\hspace{0.1cm}        &   1       &\hspace{0.05cm} \cdots   & 0
\hspace{0.1cm}  \\
\noalign{\vskip-0.1cm}
 \vdots \hspace{0.1cm}& \vdots\hspace{0.1cm}    &\hspace{0.05cm} \vdots    &
\hspace{0.05cm} \ddots    & \vdots \hspace{0.1cm}    \\
\noalign{\vskip-0.1cm}
 G_N^{\ 1}  \hspace{-0.1cm}     &   G_N^{\ 2}\hspace{0.1cm}       &   0       &\hspace{0.05cm} \cdots   & 1       \hspace{0.1cm}  \\
\end{array}\right)
,   \quad\quad 
g_0={1\over\displaystyle{\mathop{\prod}_{i=1}^N\lambda_i}}
\left(
\begin{array}{ccccc}
 \lambda_1      &   0 &   0 & \cdots   & 0    \\
 0      &   \lambda_2       & 0  & \cdots   & 0   \\
 0      &   0       &   \lambda_3       & \cdots   & 0 \\
 \vdots & \vdots    & \vdots    & \ddots    & \vdots     \\
 0      &   0       &   0       & \cdots   & \lambda_N        \\
\end{array}\right).    \label{para4} 
\end{eqnarray}
The gauge-fixed transformation is given by 
\begin{eqnarray}
g({\cal G})\longrightarrow e^{\epsilon\cdot T}g({\cal G}) U_R^{-1}=g({\cal G}'),  \label{global4} 
\end{eqnarray}
with $g({\cal G})=g_L(G)g_0(\lambda)$ and 
\begin{eqnarray}
e^{\epsilon\cdot T}\in SL(N),\quad\quad\quad
U_R=\left(
\begin{array}{ccccc}
 1      & u_{1}^{\  2} &\hspace{-0.1cm} u_{1}^{\  3} & \cdots   & u_{1}^{\  N}  \\
 0      &   1       &\hspace{-0.1cm} u_{2}^{\ 3} & \cdots   & u_{2}^{\  N}  \\
 0      &   0       &\hspace{-0.1cm}   1       & \cdots   & u_{3}^{\  N}  \\
 \vdots & \vdots    &\hspace{-0.1cm} \vdots    & \ddots    & \vdots     \\
 0      &   0       &\hspace{-0.1cm}   0       & \cdots   & 1         \\
\end{array}\right)\equiv 1+u.    \label{uu1}
\end{eqnarray}
Under this  the currents (\ref{redcurr}) transform as 
\begin{eqnarray}
 \delta J_{+i}^{\ \ j}=-{\rm Tr} (g^{-1}\partial_+g [u,T_{Ri}^{\ \ j}])
  \label{deltaJ}
\end{eqnarray}
by (\ref{property}).  They are not automatically vanishing. The reducible cases  are diffrent from the irreducible ones at this point, so that we need a care. 
The currents $J_{+i}^{\  \ i-1}$ take the form 
\begin{eqnarray}
J_{+i}^{\  \ i-1}=\left(
\begin{array}{ccccc}
 0      &\   0       & 0     &\ \hspace{-0.1cm}\cdots   &\ 0   \\
\noalign{\vskip-0.05cm}
 *     &\   0       &  0    &\ \hspace{-0.1cm} \cdots   &\ 0 \\
\noalign{\vskip-0.05cm}
 0      &\   *       &  0       &\ \hspace{-0.1cm} \cdots        &\ 0  \\
\noalign{\vskip-0.05cm}
 \vdots &\ \ddots    & \ddots    &\ \hspace{-0.1cm}\ddots    &\ \vdots     \\
\noalign{\vskip-0.05cm}
 0      &\   \cdots       &  0       &\ \hspace{-0.1cm}*  &\ 0        
\end{array}\right), \quad\quad i=2,3,\cdots, N.  \nonumber
\end{eqnarray}
The transformations (\ref{deltaJ}) become  for these currents
\begin{eqnarray}
\delta J_{+i}^{\  \ i-1}=-{\rm Tr} (g^{-1}\partial_+g [u,T_{Ri}^{\ \ i-1}]) 
 =-{\rm Tr} \left[g^{-1}\partial_+g 
\left(
\begin{array}{ccccc}
 0      &\ \hspace{0.1cm}   0       &\hspace{0.25cm} *     &\ \cdots   &\ \ *   \\
\noalign{\vskip-0.15cm}
 0     &\ \hspace{0.1cm}  0       &\hspace{0.25cm}  0    &\ \ddots   &\ \ \vdots \\
\noalign{\vskip-0.15cm}
 0      &\ \hspace{0.1cm}  0       &\hspace{0.25cm}  0       &\  \ddots       &\ \ *  \\
\noalign{\vskip-0.15cm}
 \vdots &\ \hspace{0.1cm}\vdots    &\hspace{0.25cm} \vdots    &\ \ddots    &\ \ 0     \\
 0      &\ \hspace{0.1cm}  0       &\hspace{0.25cm}  0       &\ 0   &\ \ 0         \\
\end{array}\right) \right]. \label{deltaJJ}
\end{eqnarray}
Let us impose  constraints  such as 
\begin{eqnarray}
  J_{+i}^{\  \ i-1}={\rm const.}.    \nonumber
\end{eqnarray}
Then (\ref{deltaJJ}) implies that all other currents should be constrained to be zero. Now we look for the generator $T^{U(1)}$ to define the modeified energy-momentum tensor $T_{++}$ of the form (\ref{modT}). As was shown by (\ref{ET}),  $T_{++}$ is invariant by the transformation (\ref{global4}) if $u$ satifies $[u,T^{U(1)}]=-u$. There is no solution to this equation when $u$ takes the general form given by (\ref{uu1}).  But the currents other than $J_{+i}^{\  \ i-1}$  have been constrained to be zero, so that it suffices to solve the equation assuming that $u$ to be 
\begin{eqnarray}
u=\left(
\begin{array}{ccccc}
 0      &\ \hspace{0.1cm}  *       &\hspace{0.3cm} 0     &\ \cdots   &\ \hspace{0.1cm} 0   \\
\noalign{\vskip-0.15cm}
 0      &\ \hspace{0.1cm}  0       &\hspace{0.3cm}  *    &\ \ddots   &\ \hspace{0.1cm}\vdots \\
\noalign{\vskip-0.15cm}
 0      &\ \hspace{0.1cm}  0       &\hspace{0.3cm}  0       &\  \ddots        &\ \hspace{0.1cm}0  \\
\noalign{\vskip-0.15cm}
 \vdots &\ \hspace{0.1cm}\vdots    &\hspace{0.3cm} \vdots    &\ \ddots    &\ \hspace{0.1cm}*     \\
\noalign{\vskip-0.1cm}
 0      &\ \hspace{0.1cm}  0       &\hspace{0.3cm}  0       &\ \cdots   &\ \hspace{0.1cm}0        
\end{array}\right). \nonumber
\end{eqnarray}
We find a non-trivial solution as
\begin{eqnarray}
T^{U(1)}=\left(
\begin{array}{ccccc}
\hspace{-0.1cm} \scriptstyle{N-1\over 2} &\hspace{-0.05cm} 0 &\hspace{-0.05cm} \cdots  & 0  &\hspace{-0.05cm} 0 \hspace{-0.1cm}\\
\noalign{\vskip-0.05cm}
\hspace{-0.1cm} 0  &\hspace{-0.05cm} \scriptstyle{N-3\over 2} &\hspace{-0.05cm} \cdots & 0  &\hspace{-0.05cm} 0 \hspace{-0.1cm}  \\
\noalign{\vskip-0.05cm}
\hspace{-0.1cm} \vdots  &\hspace{-0.05cm} \vdots   &\hspace{-0.05cm} \ddots &\vdots &\hspace{-0.1cm}\vdots \hspace{-0.1cm}\\
\noalign{\vskip-0.05cm}
\hspace{-0.1cm} 0 &\hspace{-0.05cm} 0 &\hspace{-0.05cm}  \cdots &\hspace{-0.15cm} \scriptstyle{-{N-3\over 2}}  &  \hspace{-0.1cm}   0 \hspace{-0.1cm}\\
\noalign{\vskip-0.05cm}
\hspace{-0.1cm}0 &\hspace{-0.05cm} 0 &\hspace{-0.05cm}\cdots &  0 &\hspace{-0.2cm}\hspace{-0.1cm} \scriptstyle{-{N-1\over 2}}\hspace{-0.1cm}  
\end{array}\right).   \label{TU1} 
\end{eqnarray}
Thus we have found the modified energy-momentum tensor $T_{++}$ which is invariant by the gauge-fixed transformation (\ref{global4}). 
 The normalization is done by choosing the quadratic Casimir to be 
$$
-C=t_{AB}T^AT^B=\sum_{i>j} (T_{Ri}^{\ \ j}T_{Lj}^{\ \ i} +T_{Lj}^{\ \ i}T_{Ri}^{\ \ j}) + \sum_{i=1}^{N-1} Q_iQ^i.
$$
$T_{Lj}^{\ \ i}$ are given similarly to $T_{Ri}^{\ \ j}$ which was given in (\ref{redcurr}), and  $Q^i$ are defined by embedding in $SL(N)$ the $U(1)$ generator of $SL(2)$ given by (\ref{T}). As the result we have $C_{adj}=-N$, or equivalently ${\rm Tr}T^AT^B={1\over 2}t^{AB}$ this time  as well. Following the procedure (\ref{G-trans})$\sim$(\ref{trans2}) or (\ref{rho})$\sim$(\ref{Killing3}) we can similarly calculate the compensators $\rho\cdot \hat H$ and $U_R^{\ -1}$,  and the Killing vectors for the group variables $G_i^{\ j}$ and $\lambda_i$ in (\ref{para4}). But we do not dare to do it here. Instead we restrict ourselves to stress on a  difference of the constraints  between both irreducible and reducible cases. We explain it for the case of $SL(3)$ for simplicity. This is the case discussed in \cite{Be}. The gauge-fixed elements are parametrized as 
\begin{eqnarray}
g_L=\left(
\begin{array}{ccc}
1 &\hspace{-0.1cm} 0\hspace{0.1cm} & 0\hspace{0.1cm} \\
G_2^{\ 1} &\hspace{-0.1cm} 1\hspace{0.1cm} & 0\hspace{0.1cm} \\
G_3^{\ 1} &\hspace{-0.1cm} G_3^{\ 2}\hspace{0.1cm} & 1 \hspace{0.1cm}
\end{array}\right), \quad\quad 
g_0=\left(
\begin{array}{ccc}
\hspace{0.2cm}{1\over \lambda\mu}\hspace{0.2cm} & 0\hspace{0.3cm} & 0 \hspace{0.2cm}\\
\hspace{0.2cm}0\hspace{0.2cm} & \lambda\hspace{0.3cm} & 0\hspace{0.2cm} \\
\hspace{0.2cm}0\hspace{0.2cm} & 0\hspace{0.3cm} & \mu \hspace{0.2cm}
\end{array}\right). \nonumber
\end{eqnarray}
According  to the above arguments the constraints for this case read 
\begin{eqnarray}
{1\over \lambda^2\mu}\partial_+G_2^{\ 1}={\rm const.}, \quad\quad
{\lambda\over\mu}\partial_+G_3^{\ 2}={\rm const.}, \quad\quad
\partial_+G_3^{\ 1}=G_3^{\ 2}\partial_+G_2^{\ 1}.  \nonumber
\end{eqnarray}
$G_2^{\ 1},G_3^{\ 2}, G_3^{\ 1}$ are the coordinates of the coset space $SL(3)/U(1)^2$, while $\lambda$ and $\mu$ are auxiliary ones. The last constraint looks like reducing the  symmetry realized by the coset space $SL(3)/U(1)^2$ to a smaller one. But no symmetry reduction occurs. The reader may check directly that the last constraint does not change the form by the Killing vectors $\delta^A G_2^{\ 1},\delta^A G_3^{\ 2}, \delta^A G_3^{\ 1}$.  This is merely a consequence of the fact that the remaining symmetry is a symmetry of the constrained currents for the reducible case as well. Owing to this fact 
 we can establish the Poisson brackets in the same way as in the previous sections.

By using the Poisson brackets (\ref{Poisson}) the Virasoro algebra can be examined  by calculating $\displaystyle{\{T_{++}(x)\mathop{,}^\otimes T_{++}(y)\}}$, in which the $U(1)$ generator  is now replaced by $T^{U(1)}$ given by (\ref{TU1}). The whole calculations go through following the similar procedure. The only argument which one might wonder about is the one given  just after the calculation (\ref{TT2}), namely, 
  $\partial_y {\rm Tr}[(\delta^Bgg^{-1}) T^{U(1)}]$  makes no  contraction with ${\rm Tr}[(\delta^Agg^{-1}) T^{U(1)}]$ and ${\rm Tr}[(\delta^Agg^{-1})(\partial_xgg^{-1}) ]$. It is correct for this case as well because the latter quntities  indeed do not contain components along the variations $\epsilon_R$s, as can be seen by calculating $u$ of (\ref{uu1}) recursively. 
 Going through this step we are led to  the Virasoro algebra 
\begin{eqnarray}
{1\over 2\pi}\int dx\eta(x) \{T_{++}(x) \mathop{,}^\otimes T_{++}(y)\}=
\eta(y)\partial_yT_{++}(y)+2\Big(\partial_y\eta(y)\Big)T_{++}(y) -{c\over 12}\partial_y^3\eta(y), \ \    \nonumber
\end{eqnarray}
with the central charge
$$ 
c=12k\sum_{i=0}^{N-1} ({N-1\over 2}-i)^2=kN(N+1)(N-1).
$$
The $SL(N)$ conformal primary which satisfies the classical exchange algebra (\ref{CEA})  is given by the first column vector of $[g_lg_0]_i^{\ j}$, which is 
$$
\Psi={1\over\displaystyle{\mathop{\prod}_{i=2}^{N}\lambda_i}}\left(
\begin{array}{c}
 1 \\
 G_2^{\ 1} \\
G_3^{\ 1}  \\
 \vdots    \\
G_N^{\ 1}
\end{array}\right). 
$$

Finally we show that the constrained currents $J_{+i}^{\ \ j}$ given by (\ref{redcurr}) has  conformal weight $0$, i.e.,
\begin{eqnarray}
{1\over 2\pi}\int dx\ \eta(x)\{T_{++}(x)\mathop{,}^\otimes J_{+i}^{\ \ j}(y) \}=\eta(y)\partial_yJ_{+i}^{\ \ j}(y).    \label{weight0}
\end{eqnarray}
It can be done closely following the proof for the irreducible case,  given in the last paragragh in subsection 4.2.  The crucial part of the proof was the calculation of (\ref{finalcal}). There use was made 
of the formula 
\begin{eqnarray}
{\rm Tr}[gug^{-1}(\partial_ygg^{-1})]=\partial_y{\rm Tr}[(\delta gg^{-1})T^{U(1)}].   \label{lastformula}
\end{eqnarray}
This formula was shown by using the explicit expressions  of the compensators and the Kiling vectors for the coset space $SL(M+N)/\{SL(M)\otimes SL(N)\otimes U(1)\}$. Those expressions for the coset space $SL(N)/U(1)^{N-1}$ are too complicated to use. Here we show it  formally. First of all we note that$$
\delta{\rm Tr}[(g^{-1}\partial_yg)T^{U(1)}]=-{\rm Tr}[(g^{-1}\partial_yg)[u,T^{U(1)}]]= {\rm Tr}[(g^{-1}\partial_yg)u],
$$
by (\ref{property}) and the constraints for $J_{+i}^{\ \ j}$. This variation can be calculated also by using (\ref{variation}) as
$$
\delta{\rm Tr}[(g^{-1}\partial_yg)T^{U(1)}]=\partial_y{\rm Tr}[(\delta gg^{-1})T^{U(1)}]. 
$$
Equating both variations yields the formula (\ref{lastformula}). Using it we calculate the Poisson bracket $\displaystyle{\{T_{++}(x)\mathop{,}^\otimes J_{+i}^{\ \ j}(y)\}}$ as before to find (\ref{weight0}).

\vspace{1cm}

\section{Conclusions}

The constrained $SL(2)$ WZWN model was much studied in the literature. In this paper we have given a proper account of the Poisson structure for  constrained $SL(N)$ WZWN models in general. Constraints are imposed on right-moving currents. Then they break the symmetry of $SL(N)$  to  a subgroup symmetry.  The key point of our arguments was to require the constrained currents to be invariant under transformations of the  remaining symmetry subgroup. To this end we considered gauging the ordinary WZWN model. Gauge-coupled currents were constrained to satisfy the requirement. This procedure led us to consider a variety of the breaking pattern of $SL(N)$. 
 The essence of the arguments in this paper was exposed for the constrained $SL(3)$ WZWN model, of which  relevant coset space was  $SL(3)/\{SL(2)\otimes U(1)\}$. 

The arguments were extended to the constrained $SL(M+N)$ WZWN model based on the coset space $SL(M+N)/\{SL(M)\otimes SL(N)\otimes U(1)\}$, which is still irreducible. We have succeeded in setting the Poisson brackets for all the group variables ${\cal G}^I=(G_i^{\ \beta}, {\Lambda'}_\alpha^{\ \beta}, \Lambda_i^{\ j})$. The group variables $G_i^{\ \beta}$ corresponding to the broken generators are coordinates of the coset space, while other variables are auxiliary, being  tangent vectors on the coset space. 
It is worth noting that for the calculation of the Poisson brackets we were able to treat $G_i^{\ \beta}$ as unconstrained variables, although they are constrained  as (\ref{c-currents}). This is owing to that fact that  the constrained currents $J_{+ i}^{\ \ \beta}$ are invariant by the remaining symmetry transformation (\ref{global3}), that is, {\it irreducibility} of the symmetry by the constraints $J_{+ i}^{\ \ \beta}={\rm const.}$. 
We have also discussed the constrained $SL(N)$ WZWN model based on the reducible coset space $SL(N)/ U(1)^{N-1}$. There the group variables $G_i^{\ j}$ describing the coset space are constrained among themselves, on the contrary to the irreducible case. But those constraints  do not reduce  the symmetry realized by the coset space.  All the constrained currents are invariant by the remaining symmetry transformation for the reducible case as well. Therefore 
the Poisson structure of the constrained WZWN model can be discussed in the same way as for the irreducible case. 

Owing to this fact we have established the Poisson brackets of the constrained WZWN models for both irreducible and reducible cases. They are consistent in the sense that  they satisfy  the Jacobi identities owing to the classical Yang-Baxter equation. By using the Poisson brackets we have obtained  the Virasoro algebra in the right-moving sector. We have also found the conformal primaries for the constrained WZWN model. They satisfy the classical exchange algebra (\ref{CEA}).  The last comment, which has no less importance, is that we have checked   the constrained currents   to have conformal weight 0. This fact has justified  the whole arguments of this paper.

 The conformal primaries of the form (\ref{primary3})  were discussed by the free field realization of the current algebra of the WZWN model in \cite{Ao2}. The free field approach provides a different point of view   about  quantum nature of the exchange algebra. Therefore we present  the arguments of \cite{Ao2} in Appendix A.

In recent years non-linear $\sigma$-models with $PSL(4|4)$ or  $PSL(2,2|4)$\cite{Bei,Bek2} 
 attracted much attention to study the string/QCD duality\cite{Ma}. We believe that 
 the WZWN models with $PSL(4|4)$ or $PSL(2,2|4)$ would play an important role there. The cases  with smaller supergroups such as
 $PSL(2|2)$\cite{Sch}, $G(1|1)$\cite{Cr}, $OSP(2|2)$\cite{Ao3}, $SL(2|1)$\cite{Ao4}  were studied by different methods and with different motivations. It would be desirable to  study them by the approach presented in this paper to shed a new light on the string/QCD duality.

\appendix

\vspace{2cm}

\section{Conformal primaries by the Berkovits method}

The conformal primary discussed in this paper may be studied by the free field realization of the WZWN model with the symmetry of the group $G$. 
The WZWN model was replaced by the linearized action\cite{Wa} 
\begin{eqnarray}
S \propto \int d^2z (\sum_{i\in \Delta_+} p^i\partial_- G_i -{1\over 2} \sum_{a=1}^{r} \partial_+\varphi^a\partial_-\varphi^a ),    \label{S'} 
\end{eqnarray}
in which $\Delta_+$ denotes the set of positive roots for the Lie algebra of $G$ and $r$ is its rank. $G_i$ parametrize the coset space $G/ U(1)^r$, while  $\varphi^a$ bosonize the $U(1)^r$ currents. They have conformal weights 
$$
h_{p^i}=1, \quad h_{G_i}= h_{\varphi^a}=0.
$$
In \cite{PRY}  $G$-symmetry conformal primaries were constructed on the basis of this  coset space 
 by using the action (\ref{S'}). In order to construct the primary based on  the more general coset space $G/\{S\otimes U(1)^l\}(\subset G/ U(1)^r)$ Berkovits 
 generalized  the action  as\cite{Bek} 
\begin{eqnarray}
S \propto \int d^2z (\sum_{i\in \Delta_+ \atop {\rm in}  G/S} p^i\partial_- G_i + \sum_{a=1}^{l} \beta^a\partial_-\gamma^a ).     \label{S''} 
\end{eqnarray}
Here $G_i$ parametrize the general coset space, while $\beta^a$ and $\gamma^a$ are bosonic ghosts realizing the $U(1)^l(\subseteq U(1)^r)$ currents.  Conformal weights of the fields  are 
$$
h_{p^i} =h_{\beta^a}=1,\quad h_{G_i}=h_{\gamma^a}=0.
$$
 But the construction of the $G$-symmetry conformal primaries can not be done as a simple generalization of that using the action (\ref{S'}). 
The difficulty was overcome by fermionization of the bosonic ghosts $\beta$ and $\gamma$ in \cite{Bek}. 
In \cite{Ao2}  an $SL(N)$ conformal primary based on $SL(N)/\{SL(N-1)\otimes U(1)\}$  was worked out as an example. 
The $SL(3)$ conformal primary given by (\ref{primary3}) in subection 3.3 is of this type. Here we show the construction of the $SL(3)$ conformal primary of this type  following  \cite{Ao2}, but adapting the normalization   to the one of subsection 3.1. The discussions can be straightforwardly generalized to the case of $SL(N)$.

The action (\ref{S''}) for the coset space $SL(3)/\{SL(2)\otimes U(1)\}$
 reads 
\begin{eqnarray}
S= k\int d^2x[ \sum_{i=1}^{2} p^i\partial_-G_i+\beta\partial_-\gamma ]
 \equiv k\int d^2x[  p\cdot\partial_-G+\beta\partial_-\gamma ]. \nonumber
\end{eqnarray}
 The system is quantized on the Euclidean world-sheet  by assuming the holomorphic  field OPEs
$$
p^j(x) G_i(y)={1\over k}{\delta^j_i \over x-y},
 \quad\quad\quad \beta(x)\gamma(y)={1 \over k}{1\over x-y}.
$$
Here $x(y)$ is a complexified coordinate, although we have so far used it to denote  the light-cone coordinate of  the right-moving sector.  Then we consider the following currents  
\begin{eqnarray}
J^A_+=-k[p\cdot \delta^AG+F^A\beta\gamma +{\partial K^A\over \partial G}\cdot\partial_+G]. \nonumber
\end{eqnarray}
Here  $\delta^AG_i$ are the Killing vectors of the coset space $SL(3)/\{SL(2)\otimes U(1)\}$ given by (\ref{G-trans}). $F^A$ and $G^A$ are some holomorphic functions of $G_i$.  They 
satisfy the $SL(3)$ current algebra 
\begin{eqnarray}
J^A_+(x)J^B_+(y)={f^{AB}_{\ \ \ C}J^C_+(y)\over x-y}- {1\over 2}{t^{AB}\over (x-y)^2},   \label{Lie}
\end{eqnarray}
if the holomorphic functions $F^A$ and $K^A$ are chosen in the following way,  
\begin{eqnarray}
J_{+Ri}&=&-k[-{1\over \sqrt 2}G_i G\cdot p +AG_i\beta\gamma+C\partial_+G_i],\quad   
J^{\ \ i}_{+L}=-{k\over \sqrt 2}p^i,  \nonumber \\
J_{+R3}&=& -{k\over \sqrt 2}G_2p^1, \quad J^{\ \ 3}_{+L}= -{k\over \sqrt 2}G_1p^2,  \label{J}  \\
\quad  J^Q&=& -{k\over 2}[G_1p^1-G_2p^2],\quad J^Y_+ =-k[-{\sqrt 3\over 2}G\cdot p +B\beta\gamma],    \nonumber
\end{eqnarray}  
with $A^2=-{3\over 2}, B^2=-1$ and $C=-{1\over k}{1\over \sqrt 2}$. 
We look for a $SL(3)$ conformal primary corresponding to $\Psi$ in  subsection 3.3, of which transformation (\ref{linear2})  is now written  as
\begin{eqnarray}
J^A_+(x)\Psi(y)= {1\over x-y}T^A\Psi(y).   \label{JPsi}
\end{eqnarray} 
In \cite{Ao2} it was found to  take  the form 
\begin{eqnarray}
\Psi= 
\left(
\begin{array}{c}
\gamma \\
\gamma G_1  \\ 
\gamma G_{2}  
\end{array}
\right),       \nonumber
\end{eqnarray}
which is similar to (\ref{primary3}). 
 But in order to show that this $\Psi$ indeed satisfies   (\ref{JPsi})  one needed the the Berkovits method\cite{Bek}. Namely the bosonic ghost pair had to be fermionized as $\beta=\partial\xi e^{-\varphi}$ and $\gamma=\eta e^\varphi$
with\cite{FMS} 
$$
\xi(x)\eta(y)=\eta(y)\xi(x)={1\over x-y},\quad\quad
\varphi(x)\varphi(y)=-\log(x-y).
$$
The quantity   $\beta\gamma$ in $J_{+Ri}$ and $J^Y_+$ given by (\ref{J}) was replaced by $a\xi\eta+b\partial_+\varphi$. When the parameters $a$ and $b$ are chosen as 
$$
k^2(a^2-b^2)=-1, \quad\quad  -k(a+b)B={1\over \sqrt 3},
$$
the current algebra (\ref{Lie}) as well as  the OPE (\ref{JPsi}) are shown to hold simultaneously.  The reader may refer to \cite{Ao2}  for further arguments on how it works.

The naive form of  the energy-momentum tensor for the action (\ref{S''}) is given by
$$
T_{++}=k[p\cdot\partial_+G + \beta\partial_+\gamma].
$$
It has  the OPEs 
\begin{eqnarray}
T_{++}(x)\Psi(y)&=& {\partial\Psi(y)\over x-y}   \label{TPsi2},  \\
T_{++}(x)T_{++}(y)&=&{c/2\over (x-y)^4}+ {2\over (x-y)^2}T_{++}(y)+
 {1\over x-y}\partial_y T_{++}(y), \label{TT1}
\end{eqnarray}
with $c=c_{pG}+c_{\beta\gamma}=6$. But in order to have the OPE 
\begin{eqnarray}
T_{++}(x)J^A_+(y)={J^A_+(y)\over (x-y)^2}+{\partial J^A_+(y)\over x-y}. 
\label{TJ}
\end{eqnarray}
$T_{++}$ also had been fermionized as\cite{FMS}
\begin{eqnarray}
T_{++}= kp\cdot\partial_+G -{1\over 2}(\partial_+\varphi)^2- {1\over 2}(\xi\partial_+\eta-(\partial_+\xi)\eta)+ Q\partial_+(\xi\eta-\partial_+\varphi).   \nonumber
\end{eqnarray}
Note that this replacement did not change the OPEs  given by (\ref{TPsi2}) and (\ref{TT1})  with any value of $Q$. Doing the same replacement in the currents $J_{+Ri}$ and $J^Y_+$ as before
 we can show that  (\ref{TJ}) is satisfied if  $Q$ is chosen to be ${3\over 2}$.

\end{document}